\documentclass[lettersize,journal]{IEEEtran}

\usepackage{amsmath,amsfonts,amssymb}
\usepackage{array}
\usepackage[caption=false,font=normalsize,labelfont=sf,textfont=sf]{subfig}
\usepackage{textcomp}
\usepackage{stfloats}
\usepackage{url}
\usepackage{verbatim}
\usepackage{graphicx}
\def\BibTeX{{\rm B\kern-.05em{\sc i\kern-.025em b}\kern-.08em
    T\kern-.1667em\lower.7ex\hbox{E}\kern-.125emX}}
\usepackage{balance}

\usepackage{cite}
\usepackage{amsthm} 

\graphicspath{{figures/}}
\usepackage{algorithm,algorithmic}
\usepackage{booktabs}
\usepackage{multirow}
\usepackage{makecell}
\newcolumntype{L}[1]{>{\raggedright\arraybackslash}m{#1}}
\newcolumntype{C}[1]{>{\centering\arraybackslash}m{#1}}

\usepackage{wasysym}
\usepackage{longtable}
\usepackage{pdflscape}
\usepackage{colortbl}
\usepackage{xcolor}
\usepackage[colorlinks=true,linkcolor=blue,citecolor=blue,urlcolor=blue]{hyperref}

\newcommand{\rev}[1]{{#1}}

\begin{document}

\title{A Survey of LLM-Driven Penetration Testing: Taxonomy, Co-Evolution, and Open Challenges}

\author{Zheyuan He, Jiaxun Dong, Zihao Li, Ting Chen, Gelei Deng, Feng Luo, Jinkun Ji,  Yuanlong Cao, and Xiapu Luo
\IEEEcompsocitemizethanks{
\IEEEcompsocthanksitem Zheyuan He is with University of Electronic Science and Technology of China, Chengdu, China, and The Hong Kong Polytechnic University, Hong Kong (email: ecjgvmhc@gmail.com). \protect
\IEEEcompsocthanksitem Jiaxun Dong, Zihao Li and Ting Chen are with University of Electronic Science and Technology of China, Chengdu, China (email: dong1505714914@gmail.com;\{zhli,brokendragon\}@uestc.edu.cn). \protect
\IEEEcompsocthanksitem Gelei Deng is with Nanyang Technological University, Singapore (email: gelei.deng@ntu.edu.sg). \protect
\IEEEcompsocthanksitem Feng Luo is with The Hong Kong Polytechnic University, Hong Kong (email: f-feng.luo@connect.polyu.hk). \protect
\IEEEcompsocthanksitem Jinkun Ji is with Four-dimensional Powerise Technology Development Co., Ltd, Beijing, China (email: v1ll4n.a5k@gmail.com). \protect
\IEEEcompsocthanksitem Yuanlong Cao is with Jiangxi Academy of Cyber Security, Nanchang, China (email: ylcao@jxnu.edu.cn). \protect
\IEEEcompsocthanksitem Xiapu Luo is with The Hong Kong Polytechnic University, Hong Kong (email: csxluo@comp.polyu.edu.hk). \protect
}
}







\markboth{}%
{Shell \MakeLowercase{\textit{et al.}}: A Sample Article Using IEEEtran.cls for IEEE Journals}

\maketitle

\begin{abstract} 
Agents4Pentest, an emerging class of LLM-based autonomous penetration testing systems, has become a rapidly growing area in security research. Despite this growth, the field still lacks a unified taxonomy, a systematic understanding of how agent architectures and evaluation benchmarks have co-evolved, and a clear characterization of remaining capability and reliability gaps.
This survey addresses these gaps through a systematic analysis of 81 papers between 2023 and 2026. We organize the literature into six categories: evaluation benchmarks, general-purpose systems, domain-specific frameworks, CTF-based systems, defense-oriented research, and surveys. We further trace a four-phase architectural evolution from text-only reasoning agents to agents trained with Reinforcement Learning with Verifiable Rewards (RLVR), showing that each transition is driven by a distinct capability bottleneck.
Our analysis yields several key findings. First, RLVR marks a shift in capability acquisition from imitation of expert demonstrations to reward-driven self-improvement, enabling agents to discover previously undocumented attack strategies. Second, CTF platforms have evolved from evaluation testbeds into dual-purpose infrastructure for both agent evaluation and RL training. Third, domain-specific frameworks improve efficiency through recurring specialization mechanisms, but their gains remain largely confined to narrow task classes and are difficult to compare across domains because existing evaluations rely on different benchmarks. Fourth, the field is expanding beyond offensive automation toward adversarial defense and security compliance. Across these categories, we identify three structurally linked open challenges: evaluation reliability, limited performance on multi-stage attack scenarios, and scarcity of high-quality training data. Overall, this survey provides a unified taxonomy, a principled basis for comparing Agent4Pentest systems, and a roadmap for future research on autonomous penetration-testing agents.
\end{abstract}


\section{Introduction}\label{sec:intro}

\IEEEPARstart{M}{odern} enterprise and cloud deployments increasingly follow services computing principles, exposing functionality through standardized APIs, dynamically composing services across organizational boundaries, and running on shared cloud and IoT infrastructure~\cite{murakami2012service,nan2017adaptive}.
Such composition chains authorization relationships across providers, creating privilege escalation paths that emerge only at the system level~\cite{lin2010dynamic,liu2015fine}.
As service boundaries multiply, authentication endpoints and access policies must be continuously validated against adversarial exploitation~\cite{ray2013trust}. 
Although penetration testing\footnote{We use \emph{pentest} as shorthand for penetration testing in this survey.}\cite{weidman2014penetration} has been regarded as a promising approach for securing services computing systems, the growing scale and dynamism of modern service deployments make manual testing insufficient for validating distributed security properties at modern deployment speed~\cite{takabi2010security,poolsappasit2011dynamic,liu2015two}.


Pentest serves as a structured security assessment in which experts simulate adversarial attacks to uncover exploitable vulnerabilities before malicious actors exploit them\cite{engebretson2013basics,meyers2022examining}.
A complete pentest spans multiple phases, including reconnaissance~\cite{kaur2017penetration}, vulnerability scanning~\cite{trickel2023toss}, exploitation~\cite{park2022fugio}, privilege escalation~\cite{de2024chainreactor}, lateral movement~\cite{he2023comprehensive}, and reporting~\cite{weidman2014penetration,engebretson2013basics}. 
Each phase requires specialized expertise, contextual reasoning over accumulated findings, and adaptive use of tools and techniques, making pentesting costly, skill-intensive, and difficult to scale with modern deployment cycles~\cite{meyers2022examining}. 
LLMs~\cite{naveed2025comprehensive} offer a promising way to reduce these barriers through multi-step reasoning, code generation, and broad security knowledge\cite{xu2024large}. 
Early work such as PentestGPT~\cite{deng2024pentestgpt} show that LLMs can reason about attack strategies and maintain context across phases, but their text-only interfaces still require human operators to execute commands and relay results, limiting the automation of pentest.

The \textit{LLM agent} paradigm~\cite{zhao2024expel} bridges this gap by equipping LLMs with tool invocation~\cite{jia2026autotool}, environmental feedback loops~\cite{liu2025survey}, and persistent memory~\cite{herrador2026spaiware}, transforming them from passive language models into active agent operators capable of issuing commands, interpreting real-time outputs, and planning across an entire engagement. 
Applied to penetration testing, such agents can autonomously identify attack surfaces from reconnaissance outputs, generate and execute exploits, interpret tool feedback for refining their strategies, and adapt their strategies across multiple phases without human intervention~\cite{xu2024large,wang2024executable}. We term this class of systems \textbf{Agent4Pentest}. Building on this paradigm, Agent4Pentest systems have rapidly evolved through multi-agent coordination~\cite{kong2025vulnbot,wang2025automated} and reinforcement learning with verifiable rewards (RLVR)~\cite{kong2025pentest,zhuo2025cyber}, driving a surge of research activity.

Despite this rapid growth, the Agent4Pentest research landscape remains fragmented. 
Agent4Pentest has expanded beyond general-purpose pentesting agents to include purpose-built benchmarks~\cite{zhang2025cybench,gioacchini2025autopenbench,shao2024nyu}, CTF-based platforms that serve as both attack systems and agent-training environments~\cite{zhuo2025cyber,shao2026towards,abramovichenigma}, domain-specific frameworks targeting particular scenarios or vulnerability classes~\cite{de2024chainreactor,yu2026chimera,jaswalawe2026awe}, and RL-based methods that move beyond prompt engineering~\cite{kong2025pentest}.
However, existing studies cover this landscape only partially.
Empirical studies either evaluate selected general-purpose systems on a single challenge set~\cite{peng2026hackers,happe2025benchmarking} or audit evaluation methodologies over a small sample of papers without synthesizing the systems themselves~\cite{happe2025benchmarking}. Perspective and modeling studies discuss LLM adoption barriers~\cite{happe2025surprising}, simulation-based attack-planning environments~\cite{wang2025unified}, or classical planning formulations of penetration testing~\cite{skandylas2025automated}, but they do not systematically examine LLM-agent architectures or real-world deployment. 
\emph{Consequently, prior work does not (i) capture the full scope of Agent4Pentest or provide a unified taxonomy, (ii) explain how agent architectures have co-evolved with benchmarks and training environments, or (iii) identify remaining capability and reliability gaps.}

To fill these gaps, we conduct a systematic survey of Agent4Pentest research landscape. Building on previous empirical studies, perspective reviews, and modeling surveys~\cite{peng2026hackers,happe2025benchmarking,happe2025surprising,wang2025unified,xu2024large}, we construct a unified analysis framework along three dimensions. First, we characterize the research landscape by organizing \emph{81 papers} between 2023 and 2026 into a six-category taxonomy (\S\ref{sec:methodology}), covering benchmarks, CTF-based platforms, general-purpose systems, domain-specific tools, RLVR-based training, and defense-oriented research. Second, we trace the co-evolution of agent architectures, evaluation infrastructure, and training paradigms, including four phases of architectural evolution (\S\ref{sec:taxonomy}) and the expansion of benchmarks from CTF platforms~\cite{taylor2017ctf,shao2024nyu,zhang2025cybench} to enterprise-scale vulnerability suites~\cite{gioacchini2025autopenbench,zhu2025cve}. Third, we examine open challenges by extending prior deployment-barrier analyses to evaluation reliability, complex attack-chain completion, and training-data scarcity. We apply this framework to benchmark papers (\S\ref{sec:benchmarks}), CTF-based platforms (\S\ref{sec:ctf}), general-purpose Agent4Pentest systems (\S\ref{sec:autopt}), and domain-specific frameworks (\S\ref{sec:domain}), comparing their task coverage, design choices, architectural assumptions, training paradigms, and reliability limitations. Table~\ref{tab:survey-comparison} in \S\ref{sec:taxonomy} summarizes the coverage differences between prior surveys and ours.

Across this framework, we derive several key findings. First, each phase of architectural evolution is driven by a distinct capability bottleneck, and RLVR marks a shift in capability acquisition from imitation of human demonstrations to reward-driven self-improvement capable of discovering previously undocumented strategies. Second, CTF platforms play a dual role as both evaluation environments and RL training substrates, making them central to the Agent4Pentest pipeline rather than merely testbeds. Third, system capability and evaluation infrastructure exhibit a co-evolutionary pattern: as agent architectures become more sophisticated, benchmarks expand in parallel, while this coupling also introduces reliability concerns that now shape the field’s central open problems. Fourth, current benchmarks remain dominated by binary success metrics, which obscure whether reported gains reflect genuine reasoning or adaptation to specific evaluation environments. Fifth, evaluation reliability, limited performance on multi-stage scenarios, and training-data scarcity are structurally linked and cannot be addressed independently. Finally, domain-specific frameworks improve efficiency through recurring mechanisms, including formal state encoding, domain knowledge injection, constrained action spaces, and specialized verification oracles, but their gains remain largely confined to narrow task classes.

The main contributions of this survey are as follows:

\begin{itemize}
    \item We derive a six-category taxonomy from 81 papers on Agent4Pentest between 2023 and 2026, and trace a four-phase architectural evolution showing that each transition is driven by a distinct capability bottleneck, from the execution-autonomy gap of text-only agents to the sample-efficiency bottleneck of RLVR-trained agents. 
    \item We formalize domain-specific Agent4Pentest frameworks as a distinct category, and identify four recurring specialization mechanisms across attack domains: formal state encoding, domain knowledge injection, constrained action spaces, and specialized verification oracles.
    \item We document the emerging role of CTF platforms as RL training substrates in the Agent4Pentest pipeline, and show that binary success metrics cannot distinguish genuine capability from benchmark alignment, while CTF-style testbeds tend to overestimate real-world capability.
    \item We consolidate open challenges, and show that evaluation reliability, limited performance on multi-stage attack scenarios, and training-data scarcity are structurally linked, making progress on any one challenge dependent on progress on the others.
\end{itemize}

As illustrated in the roadmap in Fig. 1, the remainder of this paper systematically explores the Agent4Pentest landscape based on our six-category taxonomy. Following our methodology (\S\ref{sec:methodology}) and taxonomy overview (\S\ref{sec:taxonomy}), we sequentially analyze evaluation benchmarks and various offensive/defensive strategies (\S\ref{sec:benchmarks}–\S\ref{sec:defense}), before concluding with a discussion on open challenges (\S\ref{sec:challenges} and \S\ref{sec:conclusion}).


\begin{figure}[t]
\label{fig:roadmap}
\centering
\includegraphics[width=\linewidth]{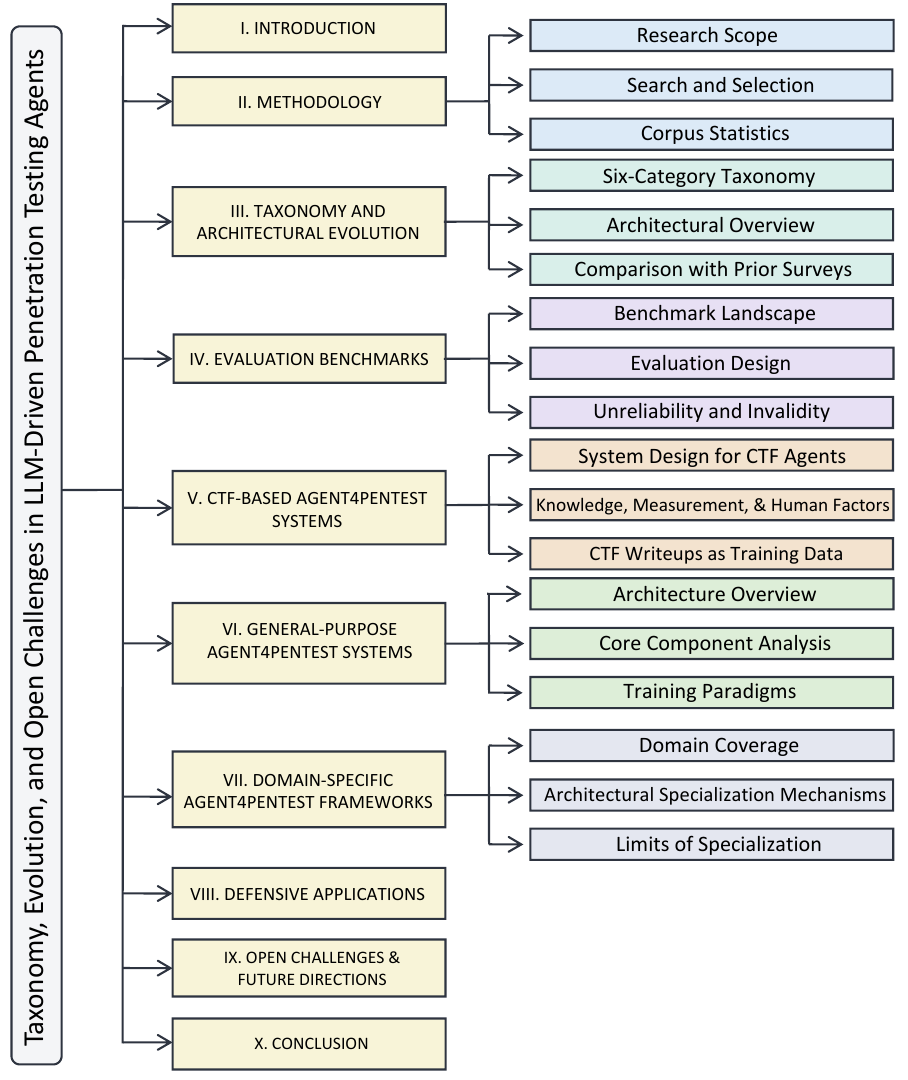}
\caption{Roadmap of this survey, showing the six-category taxonomy and the sections that cover each category.\label{fig:roadmap}}
\end{figure}

\section{Methodology}\label{sec:methodology}

\noindent\textbf{Research Scope.} 
This survey aims to systematically map the Agent4Pentest research landscape by addressing three questions: (1) what categories of systems and benchmarks have emerged in Agent4Pentest, (2) how the agent architectures have evolved over time, and (3) what open challenges cut across these categories. We follow an adapted systematic literature review protocol~\cite{keele2007guidelines}, tailored to a rapidly evolving research area driven primarily by arXiv preprints.

\noindent\textbf{Search and Selection.} 
We searched four sources: arXiv (cs.CR, cs.AI, cs.SE), ACM Digital Library, IEEE Xplore, and Google Scholar.
Search queries combined terms from two groups: (1) \textit{penetration testing, pentesting, vulnerability exploitation, CTF, capture the flag}, and (2) \textit{LLM agent, large language model, autonomous agent, reinforcement learning}.
We imposed no publication-year restriction; however, the main retrieved papers were between 2023 and June 2026, reflecting that Agent4Pentest paradigm primarily emerged in 2023.


Each candidate paper was screened against the following criteria: (i) it proposes, evaluates, or benchmarks an LLM-based agent for penetration testing or a closely related offensive security task, (ii) its full text is publicly accessible, and (iii) it contains at least one concrete technical contribution beyond a position statement.
We excluded papers that focus only on static analysis, intrusion detection, or general LLM security without a dedicated exploitation component.
We further applied forward and backward snowballing on the retained papers to recover references missed by keyword search.
The final corpus contains 81 papers, of which two are non-LLM agents retained as domain baselines: ChainReactor~\cite{de2024chainreactor}, a classical PDDL planner whose perceive--plan--act loop is architecturally analogous to LLM-based systems, and Li et al.~\cite{li2026intelligent}, a DQN-based reinforcement learning framework whose design choices inform comparison with RLVR-trained LLM agents.

\begin{figure}[t]
    \centering
    \includegraphics[width=\linewidth]{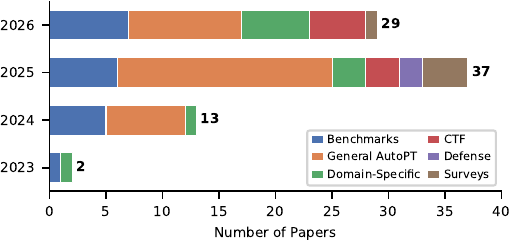}
    \caption{Papers by year and six-category taxonomy across the 81 papers.}
    \label{fig:corpus-bar}
\end{figure}

\begin{figure}[t]
    \centering
    \includegraphics[width=\linewidth]{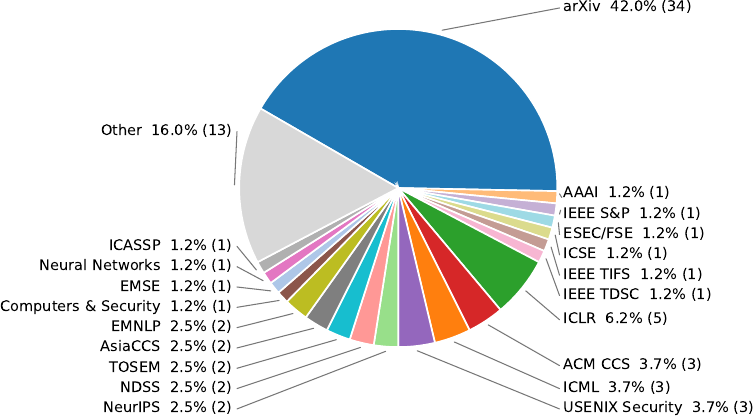}
    \caption{Distribution by publication venues across the 81 surveyed papers.}
    \label{fig:corpus-pie}
\end{figure}

\noindent\textbf{Corpus Statistics.} 
Fig.~\ref{fig:corpus-bar} and Fig.~\ref{fig:corpus-pie} summarize the distribution of the 81 papers by year, category, and publication venue.
Publication volume increased from 2 papers in 2023 to 37 in 2025 and 29 in 2026 (as of June 2026), confirming the rapid growth of Agent4Pentest research.
We label each paper using the six-category taxonomy detailed in Section~\ref{sec:taxonomy}, and Fig.~\ref{fig:corpus-bar} shows how different paper types have accumulated over time.
Specifically, general-purpose Agent4Pentest systems constitute the largest category (36 papers, 44\%), followed by evaluation benchmarks (19 papers, 23\%) and domain-specific frameworks (11 papers, 14\%). Approximately 42\% of papers appear as arXiv preprints, 17\% at top AI/ML venues (ICLR, ICML, NeurIPS, AAAI, and EMNLP), and 14\% at top security venues (ACM CCS, USENIX Security, NDSS, and IEEE S\&P)\footnote{Several papers in these venue groups appear in workshop or poster tracks rather than the main conference program.}. 


Table~\ref{tab:all-papers} provides a structured overview of all 81 papers in the corpus, organized according to the six-category taxonomy.
For Category~II papers, the Type column records the architectural type defined in \S\ref{sec:taxonomy}: text-only, tool-augmented, multi-agent, or RLVR-trained.
The MA column marks systems with explicit multi-agent coordination, while the RL column marks systems that use reinforcement learning or supervised fine-tuning.
The LLM column records the primary model family: GPT (OpenAI GPT series), OSS (open-source models, including Llama, Mistral, and Qwen), or Mix (multiple families evaluated or combined). The Env column records the evaluation environment: CTF (capture-the-flag platform), Sim (simulated Docker or VM lab), or Real (live enterprise or cloud target).

\begin{table*}[!ht]
\centering
\caption{All 81 papers organized by the six-category taxonomy.
\textbf{Type}: architectural type (Text-only / Tool-augmented / Multi-agent / RLVR-trained) for Cat.~II; benchmark type for Cat.~I; attack domain for Cat.~III.
\textbf{MA}: multi-agent coordination.
\textbf{RL}: RLVR or fine-tuning.
\textbf{LLM}: GPT = OpenAI GPT series; OSS = open-source; + = multiple families evaluated together.
\textbf{Env}: CTF = challenge platform; Sim = Docker/VM lab; Real = live target.
\label{tab:all-papers}}
\tiny
\setlength{\tabcolsep}{3pt}
\renewcommand{\arraystretch}{0.92}
\begin{tabular}{p{2.8cm} c c c p{1.8cm} c c c c c p{5.2cm} p{1.5cm}}
\hline
\textbf{Category} & \textbf{N} & \textbf{\#} & \textbf{Yr} & \textbf{Research} & \textbf{Type} & \textbf{MA} & \textbf{RL} & \textbf{LLM} & \textbf{Env} & \textbf{Key Innovation} & \textbf{Venue} \\
\specialrule{0.6pt}{0pt}{0pt}
%
\rowcolor{cyan!12}   {} & {} & 1 & 2023 & InterCode~\cite{yang2023intercode}              & CTF  & --- & --- & \rev{GPT} & CTF & Docker sandbox; multi-turn beats single-turn prompt & NeurIPS \\
\rowcolor{cyan!12}   {} & {} & 2 & 2024 & NYU CTF Bench~\cite{shao2024nyu}               & CTF  & --- & --- & GPT+Claude+Llama & CTF & 200 CSAW tasks; ranks vs.\ 1{,}176 human competitor teams & NeurIPS \\
\rowcolor{cyan!12}   {} & {} & 3 & 2024 & Fang et al.~\cite{fang2024oneday}       & CVE  & --- & --- & GPT       & Sim & 15 post-training-cutoff CVEs; 1-day and 5-day eval windows & arXiv \\
\rowcolor{cyan!12}   {} & {} & 4 & 2024 & Happe et al.~\cite{happe2024got}                   & \rev{Full}  & --- & --- & GPT+Llama   & Sim & Linux privilege escalation CVE task set with baseline & arXiv \\
\rowcolor{cyan!12}   {} & {} & 5 & 2024 & HackSynth~\cite{muzsai2024hacksynth}           & CTF  & --- & --- & GPT+Llama   & CTF & Dual-agent CTF pipeline; parametric difficulty benchmark & arXiv \\
\rowcolor{cyan!12}   {} & {} & 6 & 2024 & Shao et al.~\cite{shao2024empirical}        & CTF  & --- & --- & GPT       & CTF & GPT-4 places top 11.5\% in live CSAW competition & arXiv \\
\rowcolor{cyan!12}   {} & {} & 7 & 2024 & 3CB~\cite{anurincatastrophic}                  & \rev{Att\&CK}  & --- & --- & GPT+Claude+Llama & CTF & Three-category CTF set; context-length sensitivity study & \rev{AAAI DataSafe} \\
\rowcolor{cyan!12}   {} & {} & 8 & 2025 & Cybench~\cite{zhang2025cybench}                & CTF  & --- & --- & GPT+Claude+Llama & CTF & 40 tasks; FST difficulty proxy; 747$\times$ scale variation & ICLR \\
\rowcolor{cyan!12}   {} & {} & 9 & 2025 & AutoPenBench~\cite{gioacchini2025autopenbench} & CVE  & --- & --- & \rev{GPT} & Sim & 22 synthetic + 11 real CVEs; milestone partial-credit scoring & EMNLP \\
\rowcolor{cyan!12}   \tiny\textbf{I. Evaluation Benchmarks} & \tiny 19 & 10 & 2025 & CVE-Bench~\cite{zhu2025cve}                   & CVE  & --- & --- & \rev{GPT+Llama} & Sim & 40 real web-application CVEs; best agent solves 12.5\% & ICML \\
\rowcolor{cyan!12}   {} & {} & 11 & 2025 & APT-LLM~\cite{isozaki2025towards}             & Full & --- & --- & \rev{GPT+Llama}   & Sim & Multi-phase APT-style end-to-end engagement benchmark & UMAP \\
\rowcolor{cyan!12}   {} & {} & 12 & 2025 & PentestEval~\cite{yang2025pentesteval}        & Full & --- & --- & \rev{GPT+Claude}   & Sim & 346 stage-level tasks; mean 0.41; decision-making stage weakest & arXiv \\
\rowcolor{cyan!12}   {} & {} & 13 & 2025 & CyberGym~\cite{wang2025cybergym}              & CVE  & --- & --- & \rev{GPT+Claude+Qwen} & Sim & 1{,}507 OSS CVEs; 4-level difficulty; best 17.9\% at Level 1 & arXiv \\
\rowcolor{cyan!12}   {} & {} & 14 & 2025 & HackWorld~\cite{ren2025hackworld}             & CTF  & --- & --- & \rev{Claude+Qwen}   & Sim & Narrative-driven tasks across heterogeneous environments & \rev{ICLR} \\
\rowcolor{cyan!12}   {} & {} & 15 & 2025 & PACEbench~\cite{liu2025pacebench}             & Full & --- & --- & \rev{GPT+Claude+Gemini} & Sim & WAF + multi-host extension; all agents score 0 on WAF tasks & arXiv \\
\rowcolor{cyan!12}   {} & {} & 16 & 2026 & CTFusion~\cite{lee2026ctfusion}               & CTF  & --- & --- & \rev{GPT+Claude+Gemini}   & CTF & 71 memorization events detected; 29\% success drop after removal & arXiv \\
\rowcolor{cyan!12}   {} & {} & 17 & 2026 & Erdem et al.~\cite{erdem2026reliable}        & Full & --- & --- & GPT+Claude+Llama & Sim & 25--85\% success rate variance measured over 100 identical runs & arXiv \\
\rowcolor{cyan!12}   {} & {} & 18 & 2026 & CyberGym-E2E~\cite{shi2026cybergym}           & Full & --- & --- & GPT+Claude+Llama & Sim & 920 end-to-end tasks: vulnerability discovery + PoC + patch & \rev{ICML} \\
\rowcolor{cyan!12}   {} & {} & 19 & 2026 & ExploitGym~\cite{wang2026exploitgym}          & CVE  & --- & --- & \rev{GPT+Claude} & Sim & 898 instances spanning user-space, V8, and kernel surfaces & arXiv \\
\noalign{\hrule height 0.6pt}
\rowcolor{blue!12}  {} & {} & 20 & 2024 & PentestGPT~\cite{deng2024pentestgpt}         & \rev{Tool-augmented}   & ---        & --- & GPT     & Sim  & Pentesting Task Tree; +58.6\% sub-task completion vs.\ GPT-4 & USENIX Sec. \\
\rowcolor{blue!12}  {} & {} & 21 & 2023 & PenHeal~\cite{huang2023penheal}               & Tool-augmented  & ---        & --- & GPT     & Sim  & Two-stage pentest with automatic remediation pipeline & AutoCyber WS \\
\rowcolor{blue!12}  {} & {} & 22 & 2024 & AutoAttacker~\cite{xu2024autoattacker}        & Tool-augmented  & ---        & --- & GPT     & Sim  & MITRE ATT\&CK guided attack-chain generation & arXiv \\
\rowcolor{blue!12}  {} & {} & 23 & 2024 & Pentest-AI~\cite{bianou2024pentest}           & \rev{Multi-agent}  & \rev{\checkmark }        & --- & GPT     & Sim  & MITRE ATT\&CK multi-agent task-sequence orchestration & IEEE CSR \\
\rowcolor{blue!12}  {} & {} & 24 & 2025 & PentestAgent~\cite{shen2025pentestagent}      & \rev{Multi-agent}  & \rev{\checkmark}        & --- & GPT     & Sim  & Online CVE retrieval; 74.2\% success on 67-target benchmark & AsiaCCS \\
\rowcolor{blue!12}  {} & {} & 25 & 2025 & AutoPentester~\cite{ginige2025autopentester}  & Tool-augmented  & ---        & --- & GPT     & Sim  & RAG-augmented CVE knowledge-base retrieval for tool selection & TrustCom \\
\rowcolor{blue!12}  {} & {} & 26 & 2025 & PenTest++~\cite{al2025pentest++}              & Tool-augmented  & ---        & --- & GPT     & Sim  & Tool-priority-based exploit selection heuristic & arXiv \\
\rowcolor{blue!12}  {} & {} & 27 & 2025 & RapidPen~\cite{nakatani2025rapidpen}          & Tool-augmented  & ---        & --- & GPT     & Sim  & Search-augmented adaptive penetration plan generation & arXiv \\
\rowcolor{blue!12}  {} & {} & 28 & 2025 & Nakano et al.~\cite{nakanoguided}                 & Tool-augmented  & ---        & --- & \rev{GPT+Llama+Gemini}     & Sim  & ATT\&CK task-tree structure reduces LLM hallucinated actions & COLM \\
\rowcolor{blue!12}  {} & {} & 29 & 2025 & AutoPT-PDDL~\cite{skandylas2025automated}     & Tool-augmented  & ---        & --- & GPT     & Sim  & Labeled transition system formalization with PDDL planner & Comp.\&Sec. \\
\rowcolor{blue!12}  {} & {} & 30 & 2024 & BreachSeek~\cite{alshehri2024breachseek}      & Multi-agent & \checkmark & --- & \rev{Claude+Llama}     & Sim  & \rev{Graph-based multi-agent LLM penetration testing framework} & arXiv \\
\rowcolor{blue!12}  {} & {} & 31 & 2025 & VulnBot~\cite{kong2025vulnbot}                & Multi-agent & \checkmark & --- & \rev{GPT+Llama}     & Sim  & PTG + check-reflect loop; 19.7\% failures from tool errors & arXiv \\
\rowcolor{blue!12}  {} & {} & 32 & 2025 & CHECKMATE~\cite{wang2025automated}            & Multi-agent & \checkmark & --- & \rev{GPT+Claude}     & Sim  & PDDL symbolic planner with LLM scan-output translator & arXiv \\
\rowcolor{blue!12}  {} & {} & 33 & 2025 & Controller~\cite{geng2025controller}          & Multi-agent & \checkmark & --- & \rev{GPT+DeepSeek}     & Sim  & Controller--executor split reduces per-agent context accumulation & TrustCom \\
\rowcolor{blue!12}  {} & {} & 34 & 2025 & TermiAgent~\cite{mai2025shell}                & Multi-agent & \checkmark & --- & GPT     & Sim  & 1{,}378 CVE Docker containers expand RCE exploit coverage & arXiv \\
\rowcolor{blue!12}  {} & {} & 35 & 2025 & RefPentester~\cite{dai2025refpentester}       & \rev{Tool-augmented} & \rev{---} & --- & GPT     & Sim  & RAG knowledge-base with stage-machine self-reflection & PST \\
\rowcolor{blue!12}  \tiny\textbf{II. General-purpose Agent4Pentest} & \tiny 34 & 36 & 2025 & PTFusion~\cite{wang2025ptfusion}              & Multi-agent & \checkmark & --- & GPT     & Sim  & \rev{ Dynamic KG;MCP for context-aware fusion} & Inf. Fusion \\
\rowcolor{blue!12}  {} & {} & 37 & 2025 & PentestMCP~\cite{zhai2025pentestmcp}          & Multi-agent & \checkmark & --- & \rev{GPT} & Sim  & Model Context Protocol exposes security tools as structured calls & arXiv \\
\rowcolor{blue!12}  {} & {} & 38 & 2025 & xOffense~\cite{luong2025xoffense}             & Multi-agent & \checkmark & \checkmark & OSS & Sim  & LoRA fine-tune Qwen3-32B on pentest data; 72.72\% on AutoPenBench & arXiv \\
\rowcolor{blue!12}  {} & {} & 39 & 2025 & CAI~\cite{mayoral2025cai}                     & Multi-agent & \checkmark & --- & \rev{Claude} & Sim  & Open-source collaborative AI penetration testing system & arXiv \\
\rowcolor{blue!12}  {} & {} & 40 & 2025 & RedTeamLLM~\cite{challita2025redteamllm}      & Multi-agent & \checkmark & --- & \rev{GPT} & Sim  & Agentic red-team AI automation framework & IFIP WS \\
\rowcolor{blue!12}  {} & {} & 41 & 2026 & Red-MIRROR~\cite{khang2026red}                & Multi-agent & \checkmark & --- & \rev{Qwen}     & Sim  & Two-level reflection with majority-vote turn-level verification & arXiv \\
\rowcolor{blue!12}  {} & {} & 42 & 2026 & Deng et al.~\cite{deng2026makes}             & Multi-agent & \checkmark & --- & \rev{GPT+Claude+Gemini} & Sim & Failure taxonomy distinguishing tool error from reasoning error & arXiv \\
\rowcolor{blue!12}  {} & {} & 43 & 2026 & Incalmo~\cite{singer2026incalmo}              & Multi-agent & \checkmark & --- & \rev{GPT+Claude+Gemini} & \rev{Sim} & Decoupled LLM planning layer + specialist executor agents & IEEE S\&P \\
\rowcolor{blue!12}  {} & {} & 44 & 2026 & Li et al.$^\ddagger$~\cite{li2026intelligent} & RL-based & ---        & \checkmark & ---  & Sim  & Knowledge graph with historical decision replay enhancement & TDSC \\
\rowcolor{blue!12}  {} & {} & 45 & 2024 & Cipher~\cite{pratama2024cipher}               & RLVR-trained  & ---        & \checkmark & OSS & CTF  & SFT on 300 HackTheBox writeups; 7B model matches Llama-3-70B & Sensors \\
\rowcolor{blue!12}  {} & {} & 46 & 2025 & Pentest-R1~\cite{kong2025pentest}             & RLVR-trained  & ---        & \checkmark & OSS & CTF  & GRPO CTF RL;\rev{improves multi-step penetration testing via RL} & arXiv \\
\rowcolor{blue!12}  {} & {} & 47 & 2026 & Pen-Strategist~\cite{ginige2026pen}           & RLVR-trained  & ---        & \checkmark & OSS & Sim  & GRPO strategy fine-tuning; +47.5\% task completion on average & arXiv \\
\rowcolor{blue!12}  {} & {} & 48 & 2026 & Penetron~\cite{kong2026intent}               & RLVR-trained  & ---        & \checkmark & OSS & \rev{Sim}  & NL to Kali SFT+GRPO; outperforms models 10$\times$ larger & ICASSP \\
\rowcolor{blue!12}  {} & {} & 49 & 2024 & Fang et al.~\cite{fang2024llm}  & Tool-augmented  & ---        & --- & GPT     & Sim  & GPT-4 exploits 15 real website vulnerability types; 73.3\% pass@5 & arXiv \\
\rowcolor{blue!12}  {} & {} & 50 & 2025 & AutoPentest~\cite{henke2025autopentest} & \rev{Multi-agent} & \rev{\checkmark}     & --- & GPT     & Sim  & OWASP-specialist hierarchy; repetition detection; post-cutoff HTB & arXiv \\
\rowcolor{blue!12}  {} & {} & 51 & 2025 & Tactic Agents~\cite{ren2025automated}        & Multi-agent & \checkmark & \checkmark & GPT+OSS & Sim & Reward-prompt RL; win rate 0.3$\to$0.9; joint attack-defense & Neural Netw. \\
\rowcolor{blue!12}  {} & {} & 52 & 2025 & Huang et al.~\cite{huang2025capabilities} & Tool-augmented  & ---        & --- & \rev{GPT+Claude+Gemini} & Sim  & 5 enhancements; adaptive planning raises AutoAttacker by +27.1\% & EMNLP \\
\rowcolor{blue!12}  {} & {} & 53 & 2026 & PenForge~\cite{huang2026penforge}         & Multi-agent & \checkmark & --- & Claude    & Sim  & Dynamic expert agent from recon; 3$\times$ T-Agent on CVE-Bench & ICSE-NIER \\
\noalign{\hrule height 0.6pt}
\rowcolor{orange!14}  {} & {} & 54 & 2024 & ChainReactor$^\dagger$~\cite{de2024chainreactor} & PrivEsc & ---        & --- & ---     & Real & PDDL chain discovery; 16 chains on real EC2/DigitalOcean & USENIX Sec. \\
\rowcolor{orange!14}  {} & {} & 55 & 2025 & Perses~\cite{weber2025perses}                 & PrivEsc & \checkmark & --- & OSS     & Sim  & Small-model ensemble; 87.5\% on FreeBSD without frontier LLM & AsiaCCS \\
\rowcolor{orange!14}  {} & {} & 56 & 2026 & PrivEx-LLM~\cite{normann2026post}             & PrivEsc & ---        & \checkmark & OSS & Sim  & 5-reward RLVR; 95.8\% priv.esc.\ at 100$\times$ lower cost & arXiv \\
\rowcolor{orange!14}  {} & {} & 57 & 2026 & Happe et al.~\cite{happe2026can}              & AD      & ---        & --- & GPT+Claude & Real & First autonomous assumed-breach AD pentest on real enterprise & TOSEM \\
\rowcolor{orange!14}  {} & {} & 58 & 2026 & CHIMERA~\cite{yu2026chimera}                  & AD      & \checkmark & --- & GPT     & Sim  & LLM employee agents; 25B-entry synthetic insider-threat dataset & NDSS \\
\rowcolor{orange!14}  {} & {} & 59 & 2025 & David et al.~\cite{david2025multi}            & Web     & \checkmark & --- & GPT     & Sim  & Structured web specialist agents per vulnerability class & arXiv \\
\rowcolor{orange!14}  \tiny\textbf{III. Domain-specific Agent4Pentest} & \tiny 13 & 60 & 2026 & AWE~\cite{jaswalawe2026awe}                   & Web     & \checkmark & --- & \rev{GPT+Claude+Gemini}     & Sim  & 3-layer orchestration; 87\% XSS and 66.7\% blind SQL injection & NDSS Wksp. \\
\rowcolor{orange!14}  {} & {} & 61 & 2025 & ARACNE~\cite{nieponice2025aracne}             & SSH     & \checkmark & --- & GPT+Llama & Real & Multi-LLM plan+exec SSH agent; 60\% success on live honeypot & arXiv \\
\rowcolor{orange!14}  {} & {} & 62 & 2026 & WiFiPenTester~\cite{al2026wifipentester}      & WiFi    & ---        & --- & GPT     & Real & Governed autonomy model for IEEE 802.11 wireless assessment & Cyber Sci. \\
\rowcolor{orange!14}  {} & {} & 63 & 2026 & Ragsdale et al.~\cite{ragsdale2026ai}              & IoT     & ---        & --- & OSS     & Sim  & RL + quantization viable for IoT; LLM exceeds memory budget & IEEE Access \\
\rowcolor{orange!14}  {} & {} & 64 & 2023 & hackingBuddyGPT~\cite{happe2023getting}   & PrivEsc & ---        & --- & GPT     & Real & First LLM priv.esc.\ via live SSH; sudo/GTFObins paths & ESEC/FSE \\
\rowcolor{orange!14}  {} & {} & 65 & 2025 & AutoPT~\cite{wu2024autopt}       & Web     & ---        & --- & GPT     & Sim  & Pentest state machine (PSM); 41\% vs.\ 22\% ReAct baseline & IEEE TIFS \\
\rowcolor{orange!14}  {} & {} & 66 & 2026 & Happe et al.~\cite{happe2026llms}      & PrivEsc & ---        & --- & GPT+Llama & Real & Reflection doubles success 33\%$\to$66\%; full guidance 83\% & EMSE \\
\noalign{\hrule height 0.6pt}
\rowcolor{violet!12}  {} & {} & 67 & 2024 & Enigma~\cite{abramovichenigma}                & CTF & ---        & --- & GPT     & CTF & Hierarchical multi-phase CTF agent pipeline & ICML \\
\rowcolor{violet!12}  {} & {} & 68 & 2025 & D-Agent~\cite{udeshi2025d}                    & CTF & \rev{\checkmark}        & --- & GPT     & CTF & Diagnostic self-correction loop for iterative CTF solving & arXiv \\
\rowcolor{violet!12}  {} & {} & 69 & 2025 & Ji et al.~\cite{ji2025measuring}            & CTF & ---        & --- & \rev{GPT+Llama} & CTF & Skill taxonomy classification and RAG-based augmentation & ACM CCS \\
\rowcolor{violet!12}  \tiny\textbf{IV. CTF-based Agent4Pentest} & \tiny 8 & 70 & 2025 & CyberRL~\cite{zhuo2025cyber}                  & CTF & ---        & \checkmark & OSS & CTF & GRPO online RL; CTF platforms as RL training substrates & NeurIPS WS \\
\rowcolor{violet!12}  {} & {} & 71 & 2026 & CTFAgent~\cite{zou2026ctfagent}               & CTF & \checkmark & --- & GPT     & CTF & Multi-agent collaborative CTF solver with shared memory & JISA \\
\rowcolor{violet!12}  {} & {} & 72 & 2026 & Shao et al.~\cite{shao2026towards}               & CTF & ---        & --- & GPT+Claude & CTF & Hyperparameter tuning with LLM-as-judge evaluation framework & AAAI \\
\rowcolor{violet!12}  {} & {} & 73 & 2026 & Schachner et al.~\cite{schachner2026can}               & CTF & ---        & --- & GPT+Claude+Llama & CTF & Analysis of LLM reasoning capability boundaries on CTF tasks & arXiv \\
\rowcolor{violet!12}  {} & {} & 74 & 2026 & Striatum~\cite{hugglestone2026striatum}       & CTF & ---        & --- & GPT+Claude & CTF & Subgoal decomposition framework for structured CTF solving & arXiv \\
\noalign{\hrule height 0.6pt}
\rowcolor{red!12}    {} & {} & 75 & 2025 & CLOAK~\cite{ayzenshteyn2025cloak}             & Def. & --- & --- & GPT     & Sim & \rev{Game-theoretic active deception defense for LLM-based attack agents} & USENIX Sec. \\
\rowcolor{red!12}    \tiny\textbf{V. Defense Research} & \tiny 2 & 76 & 2025 & Sanchez et al.~\cite{sanchez2025poster}           & Def. & --- & --- & \rev{---} & Sim & Defensive application of automated penetration testing techniques & ACM CCS \\
\noalign{\hrule height 0.6pt}
\rowcolor{gray!14}   {} & {} & 77 & 2024 & Xu et al.~\cite{xu2024large}              & Survey & --- & --- & ---     & --- & SLR spanning 6 LLM security application domains & TOSEM \\
\rowcolor{gray!14}   {} & {} & 78 & 2025 & Happe et al.~\cite{happe2025benchmarking}    & Survey & --- & --- & ---     & --- & Methodology flaws audit on sampled Agent4Pentest papers & arXiv \\
\rowcolor{gray!14}   \tiny\textbf{VI. Survey Papers} & \tiny 5 & 79 & 2025 & Happe et al.~\cite{happe2025surprising}      & Survey & --- & --- & ---     & --- & 6 deployment barriers from academic, industry, black-hat sources & arXiv \\
\rowcolor{gray!14}   {} & {} & 80 & 2025 & Wang et al.~\cite{wang2025unified}            & Survey & --- & --- & ---     & --- & Simulation environment taxonomy with automated scenario generator & Comp.\&Sec. \\
\rowcolor{gray!14}   {} & {} & 81 & 2026 & Peng et al.~\cite{peng2026hackers}        & Survey & --- & --- & GPT+Claude & Sim & \rev{SoK taxonomy and evaluation of 13 AutoPT frameworks} & arXiv \\
\noalign{\hrule height 0.8pt}
\multicolumn{12}{@{}l@{}}{\parbox{\linewidth}{\scriptsize MA: multi-agent; RL: RLVR/fine-tuning; LLM: GPT=OpenAI, OSS=open-source, +=multiple families; Env: CTF=challenge platform, Sim=lab, Real=live target. $^\dagger$~Non-LLM classical planning agent; perceive--plan--act loop is architecturally analogous to LLM-based systems; included as domain baseline. $^\ddagger$~Non-LLM reinforcement learning agent (DQN); included as a general-purpose RL baseline.}}
\end{tabular}
\end{table*}

\section{Taxonomy and Architectural Evolution}\label{sec:taxonomy}

In this section, we organize the 81 papers along two dimensions: a six-category taxonomy based on primary technical contribution (\S\ref{subsec:taxonomy-cats}) and a four-phase architectural evolution that traces the field’s progression over time (\S\ref{subsec:evolution}). We then compare this survey with closely related surveys in \S\ref{subsec:comparison}.

\subsection{Six-Category Taxonomy}
\label{subsec:taxonomy-cats}
As shown in Table~\ref{tab:taxonomy}, the 81 papers in our corpus cover a broad range of research contributions, including attack systems, evaluation benchmarks, surveys, and position papers. We assign each paper to exactly one of six categories: (i) evaluation benchmarks, (ii) general-purpose Agent4Pentest systems, (iii) domain-specific frameworks, (iv) CTF-based attack systems, (v) defense research, and (vi) surveys and position papers. These categories capture the distinct roles that papers play in the Agent4Pentest research ecosystem. We derive the taxonomy through iterative open coding of the full corpus: two authors independently labeled each paper according to its primary research contribution and reconciled disagreements through discussion. Table~\ref{tab:taxonomy} summarizes the resulting distribution, and Fig.~\ref{fig:corpus-bar} shows how each category has accumulated over time.

\noindent\textbf{$\blacktriangleright$ Category I: Evaluation Benchmarks (19 papers).}
Papers in this category design controlled environments or curated task suites whose primary purpose is to measure agent capability rather than deploy new attack systems. Benchmark designs fall into three broad types: CTF challenge collections~\cite{shao2024nyu,zhang2025cybench}, real-world CVE exploitation suites that draw vulnerabilities directly from public databases~\cite{fang2024oneday,gioacchini2025autopenbench,zhu2025cve}, and controlled network environments that simulate enterprise attack scenarios. A common feature across these papers is their reliance on binary task-completion metrics.

\noindent\textbf{$\blacktriangleright$ Category II: General-purpose Agent4Pentest Systems (36 papers).}
Papers in this category propose end-to-end autonomous systems that conduct full or multi-phase penetration-testing engagements without restricting target domains or vulnerability types. We further observe a four-phase architectural evolution, from text-only prototypes~\cite{deng2024pentestgpt} to RLVR-trained agents~\cite{kong2025pentest} that surpass human-demonstration baselines on standard benchmarks. We detail this evolution  in \S\ref{subsec:evolution}.

\noindent\textbf{$\blacktriangleright$ Category III: Domain-specific Frameworks (11 papers).}
Papers in this category focus on a specific attack scenario or vulnerability class. Representative examples target Linux privilege escalation~\cite{de2024chainreactor}, enterprise Active Directory compromise~\cite{happe2026can}, and end-to-end web application testing~\cite{jaswalawe2026awe}. This restricted scope allows system architectures to be tailored to the toolchains, knowledge bases, and attack workflows of the target domain.

\noindent\textbf{$\blacktriangleright$ Category IV: CTF-based Attack Systems (8 papers).}
Papers in this category build agents for Capture the Flag competitions. CTF platforms have evolved beyond evaluation environments into reinforcement-learning substrates where agents improve by capturing real flags~\cite{zhuo2025cyber}. This dual function makes the category structurally central to the Agent4Pentest research pipeline, as analyzed in \S\ref{sec:ctf}.

\noindent\textbf{$\blacktriangleright$ Category V: Defense Research (2 papers).}
Papers in this category apply Agent4Pentest techniques for defense rather than offense, suggesting that stronger offensive agents are beginning to motivate corresponding defensive research. One line of work counters LLM attack agents by exploiting the attacking model’s own weaknesses~\cite{ayzenshteyn2025cloak}. Another embeds compliance oversight directly into the autonomous pentesting loop to keep AI agents legally accountable~\cite{sanchez2025poster}.

\noindent\textbf{$\blacktriangleright$ Category VI: Surveys and Position Papers (5 papers).}
Papers in this category synthesize or critically assess prior work without contributing new systems or benchmarks. We compare them with our survey in \S\ref{subsec:comparison}.

\begin{table}[t!]
\centering
\caption{Six-category taxonomy of the 81 surveyed papers. N denotes the paper
count and Sec.\ indicates to the section where each category is examined in depth.}
\label{tab:taxonomy}
\begingroup
\footnotesize
\setlength{\tabcolsep}{3pt}
\renewcommand{\arraystretch}{1.15}
\begin{tabular}{@{}L{0.29\linewidth}L{0.42\linewidth}C{0.07\linewidth}L{0.08\linewidth}@{}}
\hline
\textbf{Category} & \textbf{Primary contribution} & \textbf{N} & \textbf{Sec.} \\
\noalign{\hrule height 0.6pt}
I.\;\;Benchmarks       & Evaluation environments and task suites  & 19 & \S\ref{sec:benchmarks} \\
II.\;\;General AutoPT  & End-to-end autonomous pentest agents     & 36 & \S\ref{sec:autopt} \\
III.\;Domain-specific  & Specific-domain or scenario-specific tools & 11 & \S\ref{sec:domain} \\
IV.\;\;CTF Systems     & CTF agents and RL training environments  & 8  & \S\ref{sec:ctf} \\
V.\;\;Defense          & Defensive applications of agents  & 2  & \S\ref{sec:defense} \\
VI.\;Surveys           & Synthesis and position papers            & 5  & \S\ref{sec:taxonomy} \\
\hline
\textbf{Total}         &                                          & \textbf{81} & \\
\noalign{\hrule height 0.8pt}
\end{tabular}
\endgroup
\end{table}

\begin{figure}[t]
    \centering
    \includegraphics[width=\linewidth]{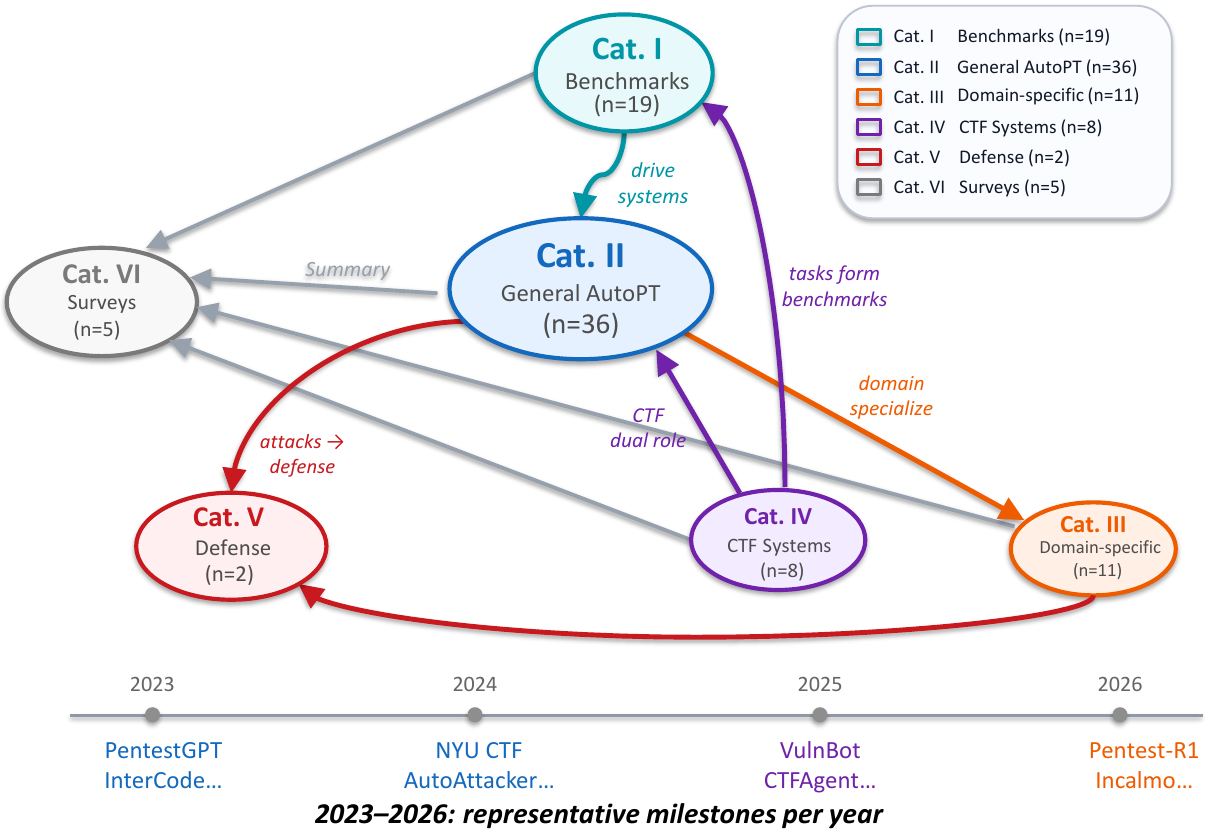}
    \caption{Research landscape of the 81 surveyed papers across six categories.
Each node represents one category, and directed edges denote structural
dependencies among categories. CTF systems (Cat. IV) provide tasks for
evaluation benchmarks (Cat. I), and also serve as evaluation environments
and RL training substrates for general-purpose AutoPT systems (Cat. II).
Benchmarks (Cat. I) drive Cat. II system development, while mature general-purpose
systems (Cat. II) further motivate domain-specific frameworks (Cat. III). The growing
capability of offensive systems (Cat. II–III) is beginning to motivate defensive
research (Cat. V). Moreover, survey papers (Cat. VI) synthesize these directions.
Timeline annotations indicate when each category first accumulated papers.
}
  \label{fig:landscape}
\end{figure}

Fig.~\ref{fig:landscape} provides a full-landscape view of the six categories, their paper counts, structural dependencies, and temporal distribution across the 2023--2026 research period.

\subsection{Architectural Overview}
\label{subsec:evolution}

\begin{figure*}[t]
\centering
\includegraphics[width=\linewidth]{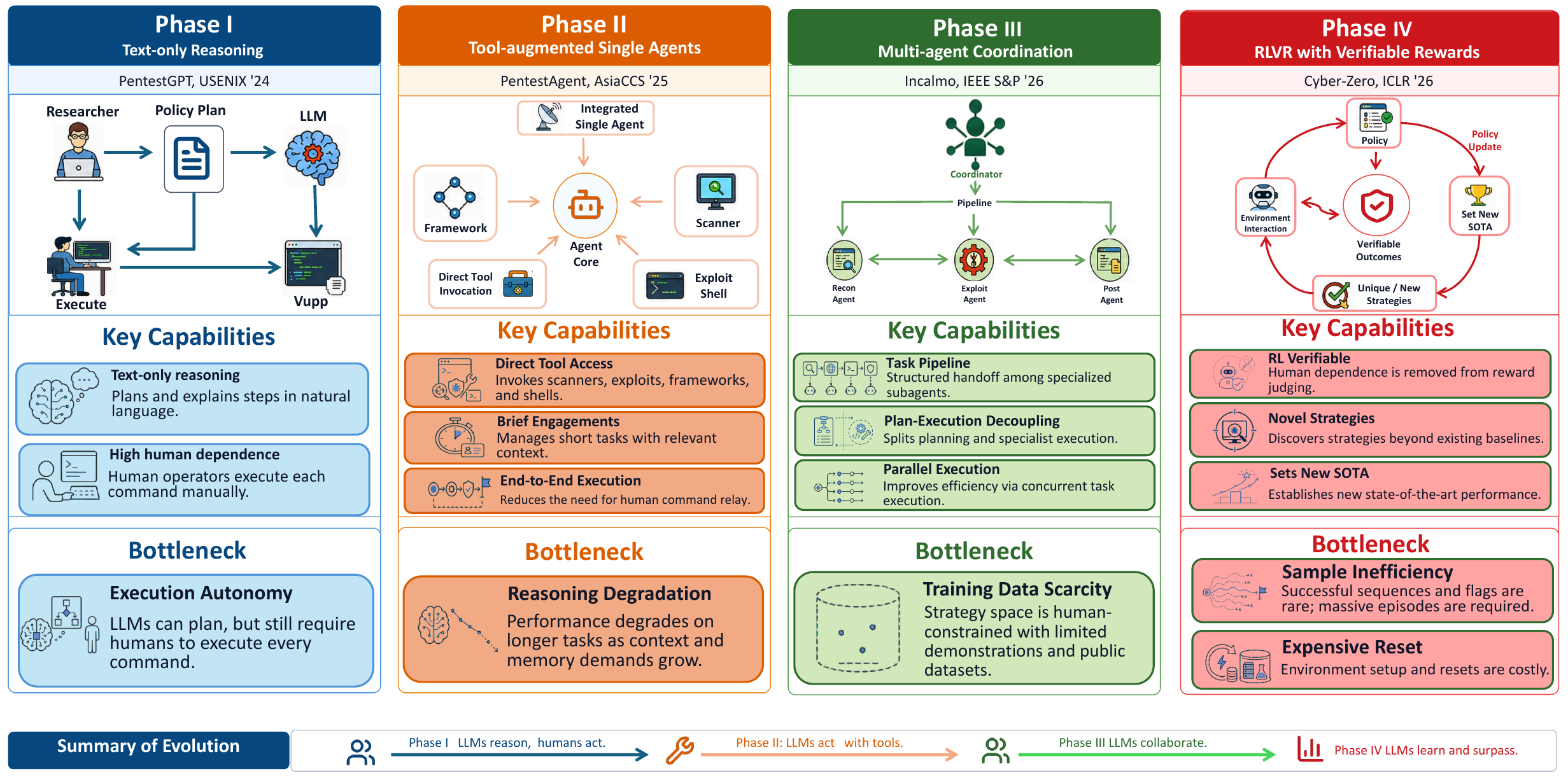}
\caption{Four-phase architectural evolution of Agent4Pentest systems.}
\label{fig:evolution}
\end{figure*}

By analyzing the 81 papers chronologically, we identify a clear four-phase progression in the design and training of Agent4Pentest systems, with each transition motivated by a capability bottleneck that the previous generation can not overcome. Fig.~\ref{fig:evolution} illustrates these phases and the bottleneck addressed by each transition.

\noindent\textbf{Phase I, Text-only Reasoning (2023).}
PentestGPT~\cite{deng2024pentestgpt} shows that a large language model can maintain a coherent attack strategy across engagement phases by operating on textual descriptions of the target state. However, Phase~I systems require a human operator to execute every command and relay terminal output back to the model, leaving \textit{execution autonomy} as the unresolved bottleneck\cite{wang2026user}.

\noindent\textbf{Phase II, Tool-augmented Single Agents (2023 to 2024).}
Subsequent work gives the LLM direct access to scanners, exploit frameworks, and shell environments, removing the need for human relay~\cite{xu2024autoattacker,shen2025pentestagent}. These systems achieve meaningful capability gains on short engagements, but long engagements cause context to accumulate until reasoning degrades. Thus, \textit{context management} becomes the unresolved bottleneck.

\noindent\textbf{Phase III, Multi-agent Coordination (2024 to 2025).}
Phase~III systems decompose the attack pipeline among specialized subagents, typically including a planner, a reconnaissance agent, an exploitation agent, and a post-exploitation agent coordinated by an orchestrator\cite{kong2025vulnbot,skandylas2025automated}. 
Distributing state across agents reduces per-agent context load and enables parallel execution. 
The limiting factor then shifts to \textit{training-data scarcity}: since these agents learn from human demonstrations, their learnable strategy space remains constrained to techniques already known to human testers.

\noindent\textbf{Phase IV, Reinforcement Learning with Verifiable Rewards (2025 to 2026).}
The latest systems reduce dependence on human demonstrations by training with reward signals derived from verifiable outcomes, such as flag capture or root-privilege acquisition~\cite{kong2025pentest,zhuo2025cyber}. 
Agents trained under this paradigm can discover attack strategies outside the human-demonstration corpus and achieve state-of-the-art results on several established benchmarks. The remaining bottleneck is \textit{sample efficiency}: successful attack sequences are rare in practice, so agents require considerable training episodes to learn reliable exploitation, and resetting the target environment for each episode is costly.

\subsection{Comparison with Prior Surveys}\label{subsec:comparison}

Table~\ref{tab:survey-comparison} compares our survey with five closely
related surveys across six coverage dimensions.
Peng et al.~\cite{peng2026hackers} conduct a large-scale empirical comparison
of 15 AutoPT frameworks, are the only related study to address the RLVR training paradigm, but their coverage of benchmarks, CTF platforms, and domain-specific tools is partial, and defense research is outside their scope.
Happe and Cito~\cite{happe2025benchmarking} provide the most thorough analysis of evaluation methodology by auditing testbed design and metrics across 19 papers, but do not cover domain-specific tools, RLVR, or defense research.
Happe~\cite{happe2025surprising} offers a broad qualitative review of systems and deployment barriers, but does not provide a systematic taxonomy or cover the RLVR paradigm. 
Wang and Cao~\cite{wang2025unified} formalize simulation-based attack planning through four representative case studies, but do not examine LLM-agent architectures, CTF platforms, or the RLVR training paradigm.
Xu et al.~\cite{xu2024large} survey the broadest literature on cyber security, covering 185 papers, but treat Agent4Pentest as a minor subtopic within a six-domain LLM-security landscape.

In summary, no prior work covers all six dimensions simultaneously. In contrast, our survey fills this gap by synthesizing a dedicated corpus of 81 papers and providing systematic coverage of CTF platforms as RL training environments, domain-specific pentesting frameworks, and defense research.

\begin{table}[t!]
\centering
\caption{Coverage comparison with related surveys.
\CIRCLE~= fully covered, \LEFTcircle~= partially covered, \Circle~= not covered.}
\label{tab:survey-comparison}
\resizebox{0.99\linewidth}{!}{
\begin{tabular}{@{}lccccccr@{}}
\toprule
\textbf{Work} & \textbf{Sys.} & \textbf{Bench.} & \textbf{CTF}
              & \textbf{Dom.} & \textbf{RLVR} & \textbf{Def.} & \textbf{N} \\
\midrule
Peng et al.~\cite{peng2026hackers}
  & \CIRCLE     & \LEFTcircle & \LEFTcircle & \LEFTcircle & \CIRCLE   & \Circle   & 15   \\
Happe \& Cito~\cite{happe2025benchmarking}
  & \LEFTcircle & \CIRCLE     & \LEFTcircle & \Circle     & \Circle   & \Circle   & 19   \\
Happe~\cite{happe2025surprising}
  & \CIRCLE     & \LEFTcircle & \LEFTcircle & \LEFTcircle & \Circle   & \LEFTcircle & 12  \\
Wang \& Cao~\cite{wang2025unified}
  & \LEFTcircle & \LEFTcircle & \Circle     & \LEFTcircle & \Circle   & \Circle   & 4    \\
Xu et al.~\cite{xu2024large}
  & \LEFTcircle & \LEFTcircle & \Circle     & \Circle     & \Circle   & \Circle   & 185  \\
\midrule
\textbf{Our work}
  & \CIRCLE   & \CIRCLE   & \CIRCLE   & \CIRCLE   & \CIRCLE   & \CIRCLE   & \textbf{81} \\
\bottomrule
\end{tabular}
}
\par\smallskip
{\footnotesize
\textit{Sys.} = general-purpose systems (Cat.~II),
\textit{Bench.} = evaluation benchmarks (Cat.~I),
\textit{CTF} = CTF-based systems (Cat.~IV),
\textit{Dom.} = domain-specific tools (Cat.~III),
\textit{RLVR} = RL with verifiable rewards,
\textit{Def.} = defense research (Cat.~V),
\textit{N} = approximate corpus size.
}
\end{table}

\section{Evaluation Benchmarks}\label{sec:benchmarks}

In this section, we survey the evaluation-benchmark papers in Category~I of our taxonomy, which measure the capabilities of Agent4Pentest systems.
Table\ref{tab:benchmarks} summarizes all 19 benchmarks. We analyze this corpus along three dimensions: task coverage (\S\ref{subsec:bench-types}), design choices for performance measurement (\S\ref{subsec:bench-design}), and evidence for measurement validity (\S\ref{subsec:bench-reliability}).

\subsection{Benchmark Landscape}\label{subsec:bench-types}

We find that Agent4Pentest benchmarks proposed in these 19 papers fall into three types: (i) CTF challenge collections, (ii) CVE exploitation suites, and (iii) full-scope scenario-based environments that simulate multi-host or multi-stage engagements. 
Early benchmarks relied primarily on isolated CTF competition tasks, whereas later work has progressively shifted toward CVE exploitation suites and full-scope multi-host scenarios that more closely reflect real pentest conditions.


\noindent\textbf{CTF challenge collections.}
Several early benchmarks~\cite{yang2023intercode,shao2024nyu,zhang2025cybench,muzsai2024hacksynth} use CTF competition tasks as the evaluation substrate for Agent4Pentest systems. 
InterCode~\cite{yang2023intercode} introduces a Docker-based interactive evaluation framework that subsequent security benchmarks build upon, showing that multi-turn interaction substantially outperforms single-turn generation and extending the framework to an initial CTF environment.
NYU CTF Bench~\cite{shao2024nyu} scales this framework to 200 CSAW competition~\cite{chung2014learning} problems across six vulnerability categories, and evaluates agents against the ranking distribution of 1,176 human competitor teams, finding that Claude 3~\cite{templeton2026scaling} ranks above the 50th percentile of human participants.
Cybench~\cite{zhang2025cybench} raises task difficulty by drawing 40 tasks from professional international competitions and introduces First-Solve Time (FST)\cite{murray2025mapping} as an objective difficulty proxy, revealing a 747-fold variation across tasks and finding that no evaluated model solves any problem whose FST exceeds eleven minutes.
HackSynth\cite{muzsai2024hacksynth} and an empirical evaluation against live CSAW competitors~\cite{shao2024empirical,wee2016self} further populate this category, with the latter reporting that GPT-4 ranks in the top 11.5\% of human teams on cryptography and binary exploitation tasks.

These results appear promising, but they rely on the assumption that competition datasets are free of memorization artifacts. A direct comparison of the same agent on a static benchmark and live competitions identifies 71 memorization events in which agents retrieve flags from training data rather than solving the challenge~\cite{lee2026ctfusion}. 
After these events are excluded, the mean success rate drops by 29\%. Although CTF platforms enable large-scale evaluation through automated flag capture, they typically present a single pre-identified target with no active defenses, unlike real engagements where attackers must first locate vulnerable hosts and operate against live controls. As a result, CTF-based benchmarks may systematically overestimate performance in real-world engagements.

\noindent\textbf{CVE exploitation suites.}
A second line of work~\cite{fang2024oneday,gioacchini2025autopenbench,zhu2025cve,wang2026exploitgym} addresses the realism gap by grounding evaluation in documented real-world vulnerabilities rather than CTF artifacts.
Fang et al.~\cite{fang2024oneday} construct 15 one-day vulnerability scenarios whose public disclosures postdate the training cutoff of the evaluated models. GPT-4 achieves an 87\% success rate when given the CVE description but only 7\% without it, indicating that vulnerability discovery remains harder than exploitation.
AutoPenBench~\cite{gioacchini2025autopenbench} combines 22 synthetic tasks with 11 real CVE scenarios covering critical vulnerabilities such as Spring4Shell~\cite{yan2024automated} and Heartbleed~\cite{durumeric2014matter}, and introduces milestone-based scoring to credit partial progress along the exploitation chain. CVE-Bench~\cite{zhu2025cve} covers 40 real web-application CVEs across eight attack objectives, and reports that the best-performing agent exploits at most 12.5\% of tasks in the one-day setting. It identifies insufficient exploration as the dominant failure mode in 55–80\% of unsuccessful cases.
ExploitGym~\cite{wang2026exploitgym} extends exploitation evaluation to 898 instances across user-space binaries~\cite{lipp2020meltdown}, V8 browser-engine targets~\cite{pantelaios2024fv8}, and Linux kernel attack surfaces~\cite{kurmus2013attack}, finding that frontier models achieve non-trivial exploitation rates even when standard mitigations such as ASLR~\cite{gruss2016prefetch} and stack canaries~\cite{dang2015performance} are enabled. Taken together, these results confirm that, even with documented CVE descriptions, current agents exploit fewer than 15\% of real vulnerabilities.

CVE-based benchmarks improve on CTF testbeds by grounding evaluation in documented real-world threats, but they still present individual prepackaged vulnerabilities rather than requiring agents to discover and chain exploits across live systems. They therefore narrow the authenticity gap relative to CTF benchmarks, but leave the discovery and multi-step reasoning gaps largely unaddressed.

\noindent\textbf{Full-scope and multi-host scenarios.}
The most recent benchmarks~\cite{wang2025cybergym, shi2026cybergym, liu2025pacebench, yang2025pentesteval} expand both scale and scenario complexity, requiring agents to chain actions across multiple hosts or complete the full vulnerability lifecycle from discovery to patch generation.
CyberGym~\cite{wang2025cybergym} draws 1,507 historical vulnerability instances from 188 open-source projects, and introduces a four-level difficulty scheme, reporting that the best agent achieves 17.9\% success on Level~1 tasks.
CyberGym-E2E~\cite{shi2026cybergym} extends this infrastructure to 920 end-to-end tasks requiring vulnerability discovery, proof-of-concept generation, and patch production.
Its end-to-end success (7.6\%) is far below the patch-only baseline (82.3\%), identifying vulnerability discovery as the primary bottleneck.
PACEbench~\cite{liu2025pacebench} adds multi-host topologies and WAF~\cite{desmet2006bridging} configurations to a CVE task set, finding that all evaluated agents score zero on WAF-bypass tasks, a scenario category absent from CTF-style benchmarks.
PentestEval\cite{yang2025pentesteval} takes a complementary approach by decomposing a full engagement into 346 stage-level tasks and applying stage-appropriate metrics, reporting a mean score of 0.41 and identifying attack decision-making and exploit generation as the weakest stages.

These results demonstrate that the competence on individual steps does not necessarily compose into full-chain success, with vulnerability discovery and multi-host reasoning remaining the primary unsolved bottlenecks. Full-scope multi-host benchmarks most closely mirror real engagement conditions by combining target discovery, vulnerability chaining, and defensive controls in a single evaluation. Their limitation is the inverse of CTF testbeds: success rates below 10\% make it difficult to differentiate agent capabilities at the current state of the art. Moreover, the complexity of scenario construction limits the number of available tasks, making statistical comparisons across systems unreliable.

\begin{figure}[t]
\centering
\includegraphics[width=\linewidth]{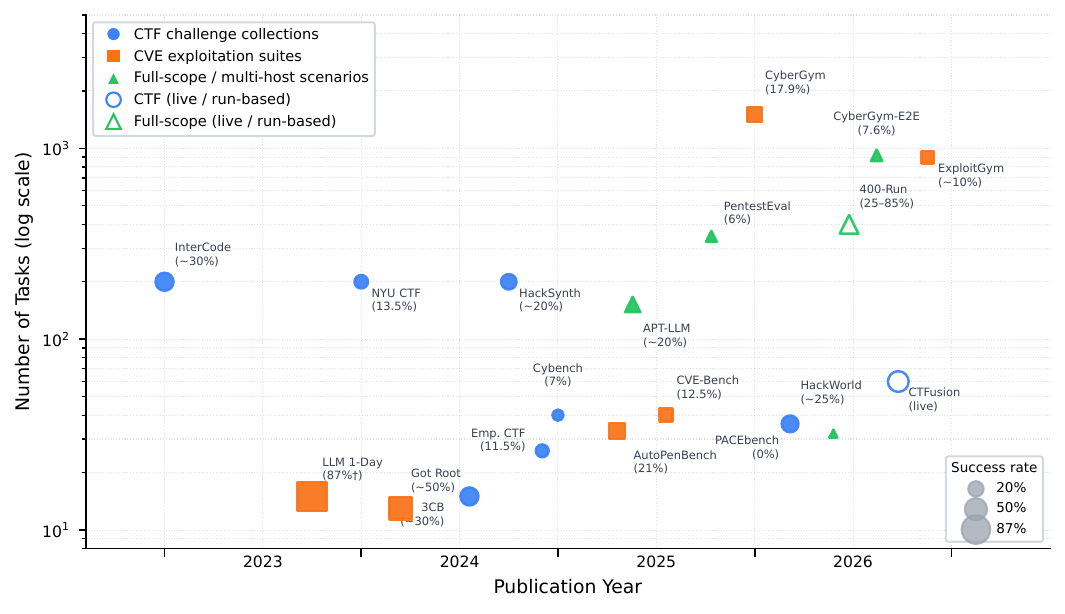}
\caption{Evolution of the 19 evaluation benchmarks (2023 - 2026). Each point represents one benchmark, with color indicating task type: CTF challenge collection, CVE exploitation suite, or full-scope scenario.
The dashed arrow highlights the shift over time toward larger, more realistic evaluations with lower reported success rates. The entry marked~\dag{} (LLM 1-Day Vulnerabilities, 87\%) reports performance when the CVE description is provided.}
\label{fig:benchmark-timeline}
\end{figure}

Fig.~\ref{fig:benchmark-timeline} displays all 19 benchmarks, with publication year on the x-axis and task count on a log-scaled y-axis. Each point is colored by task type: CTF challenge collection, CVE exploitation suite, or full-scope scenario. 
CTF benchmarks dominate in 2023–2024, whereas full-scope and CVE benchmarks become more common in 2025–2026, with the three largest task sets, CyberGym (1{,}507), CyberGym-E2E (920), and ExploitGym (898), all appearing in 2026. Averaging the best-reported success rate across all data points within each year, the per-year average falls from 30\% in 2023 to 25\% in 2024 and below 15\% in 2025–2026. The 2023 figure is based only on InterCode, the sole benchmark that year with a usable success rate, while the 2024 figure averages five benchmarks spanning all three task types.

\begin{table}[t!]
\centering
\caption{Summary of the 19 benchmark papers in Category~I, ordered by publication
date. Type: \textbf{CTF} = competition-style challenge, \textbf{CVE} = real-world
vulnerability exploitation, \textbf{Full} = full-scope, multi-host, or multi-stage
scenario.}
\label{tab:benchmarks}
\resizebox{\linewidth}{!}{%
\begin{tabular}{@{}llrcl@{}}
\hline
\textbf{Benchmark}  & \textbf{Venue}  & \textbf{Tasks}  & \textbf{Type}
                    & \textbf{Primary metric} \\
\hline
InterCode~\cite{yang2023intercode}
  & NeurIPS 2023  & $200{+}$  & CTF  & Success rate \\
NYU CTF Bench~\cite{shao2024nyu}
  & NeurIPS 2024  & 200       & CTF  & Pass@5, human rank \\
LLM 1-Day Vulns~\cite{fang2024oneday}
  & arXiv 2024    & 15        & CVE  & Pass@5 \\
Got Root~\cite{happe2024got}
  & arXiv 2024    & 13        & CVE  & Success rate \\
HackSynth~\cite{muzsai2024hacksynth}
  & arXiv 2024    & 200       & CTF  & Pass@5 \\
Emp.\ CTF Eval~\cite{shao2024empirical}
  & NeurIPS 2024  & 26        & CTF  & Pass@10, human rank \\
Cybench~\cite{zhang2025cybench}
  & ICLR 2025     & 40        & CTF  & SR, FST \\
AutoPenBench~\cite{gioacchini2025autopenbench}
  & EMNLP 2025    & 33        & CVE  & Milestone (MC/MS) \\
3CB~\cite{anurincatastrophic}
  & arXiv 2024    & 15        & CTF  & ATT\&CK completion rate \\
CVE-Bench~\cite{zhu2025cve}
  & ICML 2025     & 40        & CVE  & Success@5 \\
APT-LLM~\cite{isozaki2025towards}
  & UMAP 2025     & 152       & Full & Step success \\
PentestEval~\cite{yang2025pentesteval}
  & arXiv 2025    & 346       & Full & Stage-specific metrics \\
CyberGym~\cite{wang2025cybergym}
  & ICLR 2026     & 1{,}507   & CVE  & Pass@1/5 \\
HackWorld~\cite{ren2025hackworld}
  & ICLR 2026     & 36        & CTF  & Success rate (OCR-tolerant) \\
PACEbench~\cite{liu2025pacebench}
  & ICML 2026     & 32        & Full & Pass@5 \\
CTFusion~\cite{lee2026ctfusion}
  & arXiv 2026    & live      & CTF  & Pass@3 \\
400-Run Study~\cite{erdem2026reliable}
  & arXiv 2026    & 400 runs  & Full & Exploitation rate \\
CyberGym-E2E~\cite{shi2026cybergym}
  & ICML 2026     & 920       & Full & S1--S4 cascade \\
ExploitGym~\cite{wang2026exploitgym}
  & arXiv 2026    & 898       & CVE  & RCE success \\
\hline
\end{tabular}}
\par\smallskip
{\footnotesize
\textit{Pass@k} = success rate over $k$ independent attempts (best of $k$);
\textit{SR} = single-run success rate;
\textit{FST} = First-Solve Time (median human solve time, used as an objective difficulty proxy);
\textit{MC/MS} = command-level and phase-level milestones;
\textit{S1--S4} = four-stage cascade (PoC crash, patch prevents crash, functional tests pass, and patch matches ground truth);
\textit{RCE} = code execution verified dynamically;
\textit{ATT\&CK} = MITRE ATT\&CK tactical category completion rate;
\textit{live} = task count varies with active competition schedule.
}
\end{table}

\subsection{Evaluation Design}\label{subsec:bench-design}

Most benchmarks adopt binary task completion as the primary evaluation metric, recording only whether a task is fully solved. For example, a CTF task receives a score of 1 if the correct flag is submitted and 0 otherwise, with no credit for intermediate progress. This design lacks the granularity to distinguish strong agents from weak ones: an agent that completes every exploitation step except final flag submission receives the same zero score as one that never meaningfully attempts the task.

A smaller set of benchmarks introduce finer-grained metrics that credit partial progress along the exploitation chain. PentestEval~\cite{yang2025pentesteval} decomposes a full engagement into six sequential stages and applies stage-specific metrics, including rank correlation for attack-decision quality and functional correctness for exploit-code generation. It reports an overall mean score of 0.41 and identifies attack decision-making and exploit generation as the weakest stages, each scoring approximately 0.25. 
AutoPenBench~\cite{gioacchini2025autopenbench} introduces command-level and phase-level milestones scored by an LLM-as-a-judge component, reporting 21\% success for the autonomous agent and 64\% for the human-assisted variant. This gap highlights the difference between current automation capability and what close human guidance can achieve. Fine-grained metrics reveal stage-level bottlenecks that binary scores conceal, but they require substantially more annotation effort and reduce cross-system comparability. As a result, task-completion rate remains the dominant headline metric in practice.

\begin{figure}[t]
\centering
\includegraphics[width=\linewidth]{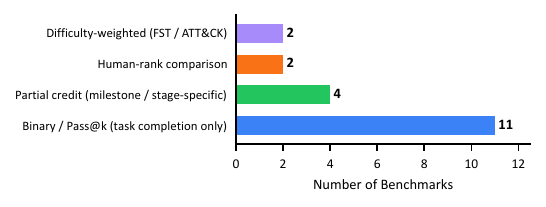}
\caption{Distribution of primary evaluation metrics across the 19 benchmarks in Category~I.
Binary pass-or-fail and pass@$k$ metrics dominate, covering 11 benchmarks including InterCode~\cite{yang2023intercode}, NYU CTF Bench~\cite{shao2024nyu}, HackSynth~\cite{muzsai2024hacksynth}, CVE-Bench~\cite{zhu2025cve}, and PACEbench~\cite{liu2025pacebench}.
Human-rank comparison is used by two benchmarks (NYU CTF Bench~\cite{shao2024nyu} and the empirical CSAW study~\cite{shao2024empirical}), which evaluate agents against the score distribution of human competitor teams.
Partial-credit metrics covering milestone or stage-specific scoring are used by four benchmarks (AutoPenBench~\cite{gioacchini2025autopenbench}, APT-LLM~\cite{isozaki2025towards}, PentestEval~\cite{yang2025pentesteval}, and CyberGym-E2E~\cite{shi2026cybergym}).
Difficulty-weighted metrics are used by two benchmarks, Cybench~\cite{zhang2025cybench} employing First-Solve Time and 3CB~\cite{anurincatastrophic} employing ATT\&CK completion rate.}
\label{fig:benchmark-metrics}
\end{figure}

A second challenge is difficulty calibration. Benchmarks dominated by easy tasks inflate aggregate success rates, making it difficult to determine whether a high-scoring agent is genuinely capable or merely benefits from the task distribution. Cybench~\cite{zhang2025cybench} reports a 747-fold variation in FST across its 40 tasks, showing that aggregate success rates can be driven by the proportion of easy problems rather than agent capability at the intended difficulty level. 
A complementary issue is evaluation variance: repeating the same attack scenario 100 times per agent yields success rates ranging from 25\% to 85\% across models~\cite{erdem2026reliable}, suggesting that single-run evaluations provide unreliable estimates of agent capability. FST offers an objective, human-derived difficulty proxy that reduces selection bias, while multi-run evaluation controls variance; however, both substantially increase evaluation cost and are rarely adopted together. Fig.~\ref{fig:benchmark-metrics} shows the distribution of primary metric types across all 19 benchmarks, confirming that binary pass-or-fail remains the dominant evaluation design despite its known limitations.

\subsection{Unreliability and Invalidity}
\label{subsec:bench-reliability}

Based on the analysis in \S\ref{subsec:bench-types} and \S\ref{subsec:bench-design}, we identify two fundamental problems that limit the reliability and validity of current benchmark results.
First, data contamination can inflate performance on static datasets, making it difficult to determine whether a high-scoring agent has genuine problem-solving capability or has memorized benchmark artifacts. Second, the structural mismatch between CTF tasks and real engagements limits external validity: success on isolated CTF challenges does not necessarily transfer to the multi-step reasoning, target discovery, exploit chaining, and defense evasion required in real deployments.

\noindent
\emph{i). Data Contamination:}
It arises when the model underlying Agent4Pentest systems has seen benchmark tasks during training, and can recall the correct answer rather than solve challenges. In this case, reported success rates measure memorization rather than reasoning ability. Lee et al. quantify this effect by comparing the same agent on a fixed static benchmark and live competitions held during evaluation window, identifying 71 memorization events and a 29\% drop in mean success rate after removing them~\cite{lee2026ctfusion}. This result suggests that published CTF benchmark scores may partially reflect memorization artifacts, and that static benchmark results should be interpreted cautiously unless contamination is ruled out.

\noindent
\emph{ii). The Structural Mismatch between CTF Tasks and Real Engagements:} In a typical CTF challenge, the target host is pre-identified and isolated, the vulnerability is known to exist, and active defenses are absent. By contrast, real engagements require agents to locate vulnerable hosts among benign ones, chain exploits across host boundaries, and operate against live defensive controls. Recent studies provide evidence for this gap: adding multi-host topologies and WAF configurations to a CVE task set reduces all evaluated agents to zero success on WAF-bypass tasks~\cite{liu2025pacebench}, while decomposing a full engagement into stages shows that, even when agents achieve an average stage-level score of 0.41, end-to-end success for fully autonomous agents remains below 6\%~\cite{yang2025pentesteval}. These results indicate that individual-step performance does not reliably predict full-chain attack capability, exposing a gap between benchmark scores and field-ready competence.

\begin{figure}[t]
\centering
  \centering
  \includegraphics[width=\linewidth]{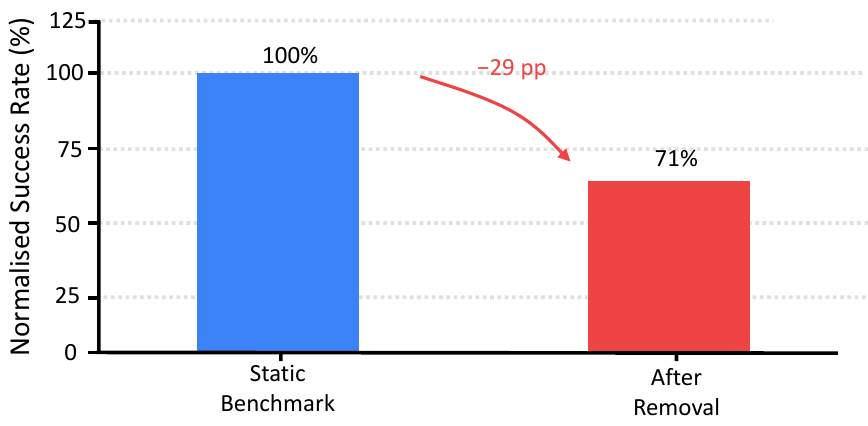}
  \caption{Contamination effect: removing 71 memorisation events detected by the CTFusion~\cite{lee2026ctfusion} live-competition comparison reduces the mean success rate by 29 percentage points relative to static benchmark results.}
  \label{fig:benchmark-contamination}
\end{figure}

\begin{figure}[t]
  \centering
  \includegraphics[width=0.88\linewidth]{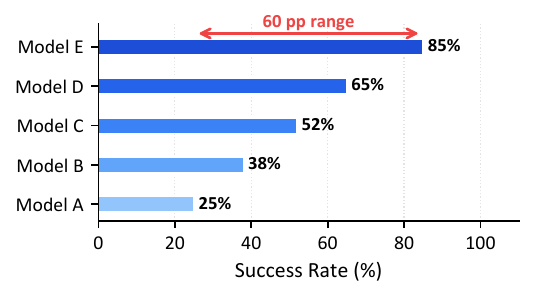}
  \caption{Evaluation variance: repeating an identical attack scenario 100 times per agent (400-run study) yields success rates spanning 25--85\% across five evaluated models, making single-run comparisons unreliable estimates of agent capability.}
  \label{fig:benchmark-variance}
\end{figure}

Fig.~\ref{fig:benchmark-contamination} and Fig.~\ref{fig:benchmark-variance} quantify these reliability concerns. Fig.~\ref{fig:benchmark-contamination} shows the CTFusion result, where removing 71 memorization events reduces the mean success rate by 29\%\cite{lee2026ctfusion}. Fig.~\ref{fig:benchmark-variance} shows a 400-run evaluation in which the same scenario, repeated 100 times per agent, yields success rates ranging from 25\% to 85\% across five agents~\cite{erdem2026reliable}.

Taken together, these findings suggest that current benchmarks may overestimate real-world Agent4Pentest capability. The field has begun to respond by moving from isolated CTF challenges (2023–2024) to CVE exploitation suites (2024–2025) and then to full-lifecycle, defense-aware evaluation frameworks (2025–2026), with each shift aiming to narrow the gap between measured and deployed performance. Addressing contamination requires live or regularly refreshed task sets, while improving external validity requires benchmarks that incorporate target discovery, multi-host topologies, exploit chaining, and active defenses. Standardizing these practices is essential for future benchmarks to provide reliable and valid measures of real-world Agent4Pentest capability.


\section{CTF-based Agent4Pentest Systems}\label{sec:ctf}

This section surveys the agent4pentest systems targeting CTF competitions in Category~IV of our taxonomy, covering system design (\S\ref{subsec:ctf-design}), knowledge and measurement findings (\S\ref{subsec:ctf-knowledge}), and agent training methods (\S\ref{subsec:ctf-training}).

\subsection{System Design for CTF Agents}\label{subsec:ctf-design}

We first describe the general design of Agent4Pentest systems for CTF problems. A CTF problem provides competitors with a problem description, access to a target environment such as a binary file or running network service, and an automated flag-verification mechanism that confirms success when the correct flag is submitted. 
To capture the flag, competitors infer the vulnerability category from the problem description, select appropriate tools, apply them to the target, and iterate on the results until the flag is extracted. A CTF agent follows the same workflow: it forms an initial hypothesis about the vulnerability type, selects and invokes security tools against the target, and refines its approach over multiple rounds until the flag is captured.

At a high level, these agents couple an LLM reasoning core with security-specific tools and a state-tracking component that records attempted actions. The LLM generates attack hypotheses, the tools execute them against the target environment, and the resulting outputs guide subsequent steps. In practice, however, implementing this loop reliably is difficult, and the design space remains large. This difficulty stems from three bottlenecks: tool access, context management, and hallucination control. Each is addressed by one of the four systems discussed in this subsection~\cite{abramovichenigma,udeshi2025d,zou2026ctfagent,hugglestone2026striatum}.

\begin{table}[t!]
\centering
\caption{Summary of four CTF agent systems in \S\ref{subsec:ctf-design}, organized by the design bottleneck each primarily addresses. ``---'' indicates the bottleneck is not the primary focus of that system.}
\label{tab:ctf-systems}

\begingroup
\normalsize
\renewcommand{\arraystretch}{1.55}
\setlength{\tabcolsep}{8pt}

\resizebox{\linewidth}{!}{%
\begin{tabular}{
@{}
C{3.1cm}
C{2.2cm}
C{3.0cm}
C{3.0cm}
C{3.2cm}
@{}
}
\hline
\textbf{System}
&
\makecell[c]{\textbf{Tool}\\\textbf{Access}}
&
\makecell[c]{\textbf{Context}\\\textbf{Management}}
&
\makecell[c]{\textbf{Hallucination}\\\textbf{Control}}
&
\makecell[c]{\textbf{Best}\\\textbf{Result}}
\\
\hline
EnIGMA~\cite{abramovichenigma}
&
Persistent IATs
&
Output summarizer
&
Soliloquizing
&
13.5\% (NYU CTF)
\\
D-CIPHER~\cite{udeshi2025d}
&
---
&
3-agent split
&
---
&
19.0\% (NYU CTF)
\\
CTFAgent~\cite{zou2026ctfagent}
&
---
&
Task tree
&
---
&
88th pct.\ (PicoCTF)
\\
STRIATUM-CTF~\cite{hugglestone2026striatum}
&
---
&
---
&
MCP validator
&
86.7\%; 1st/22 (live)
\\
\hline
\end{tabular}
}
\endgroup

\end{table}

\noindent
\emph{-- i). Interactive tool access} is a primary gap between current agent performance and human expert capability on CTF tasks. EnIGMA~\cite{abramovichenigma} addresses this gap through Interactive Agent Tools (IATs), which maintain persistent sessions with security utilities such as GDB and pwntools, while long-output summarizers condense tool responses that would otherwise exceed the context window. 
Ablation experiments show that IATs and in-context demonstrations provide independent gains of 2.1\% and 6.2\%, respectively, and the full system achieves a 13.5\% solve rate on NYU CTF Bench, more than three times the prior best. The work\cite{abramovichenigma} also identifies \textit{soliloquizing}, a failure mode in which the agent fabricates tool-output lines without issuing an actual tool call, which binary solve-rate metrics cannot detect.

\noindent
\emph{-- ii). Context management} is the second bottleneck, as long engagements cause single-agent systems to accumulate irrelevant outputs until reasoning degrades. 
D-CIPHER~\cite{udeshi2025d} addresses this issue through a three-agent decomposition: an Auto-prompter generates a task-specific initial prompt, a Planner delegates subtasks, and each Executor operates with a fresh context bounded to the five most recent messages. 
Context management is a second bottleneck, as long engagements cause single-agent systems to accumulate irrelevant outputs until reasoning degrades. D-CIPHER~\cite{udeshi2025d} addresses this issue through a three-agent decomposition: an Auto-prompter generates a task-specific initial prompt, a Planner delegates subtasks, and each Executor operates with a fresh context bounded to the five most recent messages. 
Ablating the Planner reduces the solve rate from 19.0\% to 14.0\%, confirming that structured delegation drives the gain. The full system achieves 19.0\%, 22.5\%, and 44\% solve rates on NYU CTF Bench, CyBench, and HackTheBox~\cite{hackthebox26}, respectively, under Claude 3.5 Sonnet. CTFAgent~\cite{zou2026ctfagent} takes a complementary approach, replacing bounded context windows with a stateful task tree that records the long-term attack plan independently of the active LLM context, allowing strategies to persist across interactions of arbitrary length. 
Evaluated on 89 PicoCTF problems against more than ten thousand human teams, its fully autonomous mode ranks above the 88th percentile, while its human-in-the-loop mode reaches the 94th percentile. Together, these systems show that context-management infrastructure can mitigate current context limitations, while the remaining performance gap increasingly reflects core LLM reasoning capability rather than execution context alone.

\noindent
\emph{-- iii). Hallucination control} addresses a distinct failure mode in which agents generate malformed or nonexistent commands that the execution layer cannot run. 
STRIATUM-CTF~\cite{hugglestone2026striatum} inserts a protocol layer between the LLM and the execution layer, using an MCP schema validator to restrict every proposed action to commands that the system can dispatch. 
This constraint-driven design achieves 86.7\% success across 15 CTF problems, and a prompt-length ablation shows no significant difference between a 4,147-line and a 55-line context, suggesting that reliable execution depends more on protocol constraints than on documentation volume. The system then competed in a real university CTF event and finished first among 22 teams, with a 10-point margin over the highest-scoring human team, providing direct evidence that benchmark performance can transfer to live competition.

\begin{figure}[t]
\centering
\includegraphics[width=\linewidth]{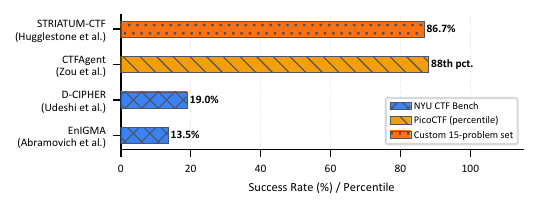}
\caption{Reported results of the four CTF agent systems, grouped by evaluation platform.
EnIGMA~\cite{abramovichenigma} and D-CIPHER~\cite{udeshi2025d} are evaluated on the shared NYU CTF Bench and are therefore directly comparable.
CTFAgent~\cite{zou2026ctfagent} reports an 88th-percentile ranking on PicoCTF, which is a percentile rather than a solve rate.
STRIATUM-CTF~\cite{hugglestone2026striatum} reports 86.7\% success on a custom 15-problem set and places first among 22 teams in a live competition. Results from different platforms use different metrics and are not directly comparable.}
\label{fig:ctf-systems}
\end{figure}

Fig.~\ref{fig:ctf-systems} shows each system’s result, colored by evaluation platform. EnIGMA~\cite{abramovichenigma} and D-CIPHER~\cite{udeshi2025d} both report results on NYU CTF Bench; D-CIPHER reaches 19.0\% compared with EnIGMA’s 13.5\%, a 5.5 percentage-point gain attributable to its three-agent context decomposition. CTFAgent~\cite{zou2026ctfagent} reports an 88th-percentile ranking on PicoCTF, which is a percentile rather than a solve rate, while STRIATUM-CTF~\cite{hugglestone2026striatum} reports 86.7\% success on a custom 15-problem set. Only the NYU CTF Bench results support direct comparison; the PicoCTF and custom-set results reflect different platforms and metrics.

The four systems~\cite{abramovichenigma,udeshi2025d,zou2026ctfagent,hugglestone2026striatum} converge on common design principles from different starting points. Interactive tool support, structured task delegation, external strategic memory, and protocol-enforced action filtering each improve performance independently, as supported by ablation studies reported in at least two of the four papers\rev{~\cite{abramovichenigma,udeshi2025d,hugglestone2026striatum}}. The fact that multiple independent groups arrived at similar solutions to the same three challenges suggests that tool access, context management, and hallucination control constitute the emerging core of CTF-agent design.

\subsection{Knowledge, Measurement, and Human Factors}\label{subsec:ctf-knowledge}

Next, we explore three observations about how CTF-agent capability is understood and measured: the gap between model knowledge and task performance, the coarseness of binary evaluation metrics, and the effect of AI assistance on human skill development. We find that current evaluation practices can overstate actual CTF capability, because knowledge scores do not imply task-solving ability, binary pass rates do not capture partial exploitation progress, and AI-assisted performance does not necessarily reflect independent skill.

Frontier models have accumulated substantial domain knowledge for CTF problems, yet this knowledge does not directly translate into task-solving capability. 
Ji et al.\cite{ji2025measuring} quantify this gap by constructing CTFKnow, a benchmark of 1,996 multiple-choice and 1,996 open-ended questions derived from more than 700 competitions. GPT-4o achieves 87.83\% accuracy in the multiple-choice setting, but performs substantially worse on open-ended items. 

\begin{figure}[t]
\centering
\includegraphics[width=\linewidth]{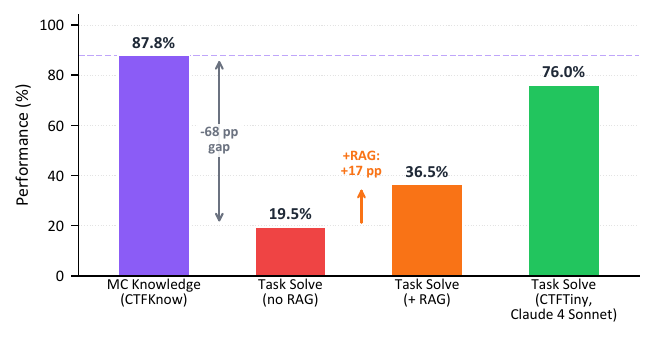}
\caption{Knowledge–performance gap on CTF tasks. GPT-4o achieves 87.83\% accuracy on multiple-choice knowledge questions in CTFKnow~\cite{ji2025measuring}, but only 19.5\% on actual CTF task solving without retrieval augmentation. Adding a RAG component increases the task solve rate to 36.5\%. Under the richer CTFTiny evaluation~\cite{shao2026towards}, Claude 4 Sonnet reaches a 76\% solve rate. These results indicate that the main bottleneck lies in procedural application of knowledge rather than knowledge storage.}
\label{fig:ctf-knowledge}
\end{figure}

Actual task-solving performance drops further, confirming that applying knowledge, rather than storing it, is the primary bottleneck. To address this gap, \rev{the study~\cite{ji2025measuring}} introduces a CTF agent that combines retrieval-augmented generation with an interactive environment component, increasing the solve count on InterCode-CTF from 39 to 73 tasks. 
Fig.\ref{fig:ctf-knowledge} plots four results side by side: GPT-4o’s multiple-choice accuracy on CTFKnow~\cite{ji2025measuring} (87.83\%), its task solve rate without RAG (19.5\%), its task solve rate with RAG (36.5\%), and Claude 4 Sonnet’s solve rate on CTFTiny~\cite{shao2026towards} (76\%). The first three results illustrate the knowledge–task gap and the RAG gain for GPT-4o; the fourth is a separate result on a different benchmark and model and is therefore not directly comparable.

Binary pass-or-fail metrics conceal partial progress's extent achieved by agents, motivating finer-grained evaluation tools. 
Shao et al. introduce CTFTiny, a 50-task benchmark stratified across four difficulty levels, and CTFJudge, an LLM-as-a-judge framework that scores agent trajectories along six dimensions and computes a composite CTF Capability Index (CCI) to credit partial attack-chain completion~\cite{shao2026towards}. 
A hyperparameter study\rev{\cite{shao2026towards}} finds that high temperature, approximately 1.0, and medium context length, 4,096 tokens, suit capable models, whereas smaller models exhibit a non-monotonic temperature response, cautioning against transferring settings across model tiers. Under Claude 4 Sonnet, CTFJudge reports a 76\% solve rate and a CCI of 77.5–84.5 on CTFTiny, providing a finer-grained view of frontier-model capability than binary metrics alone.

High performance under AI assistance does not imply that the human operator has developed independent capability. Schachner et al.~\cite{schachner2026can} examine this gap by following a novice participant for approximately one year while they use an AI framework to compete in a national CTF event. AI assistance gives the newcomer a strategic overview of the attack landscape, structured investigation steps, and reduced cognitive load, but also induces strategic passivity, in which the participant defers to the AI rather than forming an independent analysis. This finding suggests that CTF agents in collaborative settings should expose high-level reasoning, not only executable outputs, because strategic guidance provides greater benefit for learning users.

In summary, all three measurement settings risk overestimating capability by conflating performance under favorable conditions with genuine task-solving ability. High declarative knowledge does not guarantee procedural application; a nearly complete exploitation chain receives zero credit if the final flag submission fails; and strong performance under AI guidance may not transfer to independent attempts. The community should therefore move from knowledge tests toward task-solving benchmarks, from binary pass/fail metrics toward partial-credit evaluation, and from human-assisted settings toward fully autonomous evaluation, because mixing these conditions makes cross-system comparison unreliable.

\subsection{CTF Writeups as Training Data}\label{subsec:ctf-training}

Finally, we describe the most distinctive training approach in CategoryIV, in which CTF competition writeups serve as the primary source of agent training data. 

\begin{figure}[t]
\centering
\includegraphics[width=\linewidth]{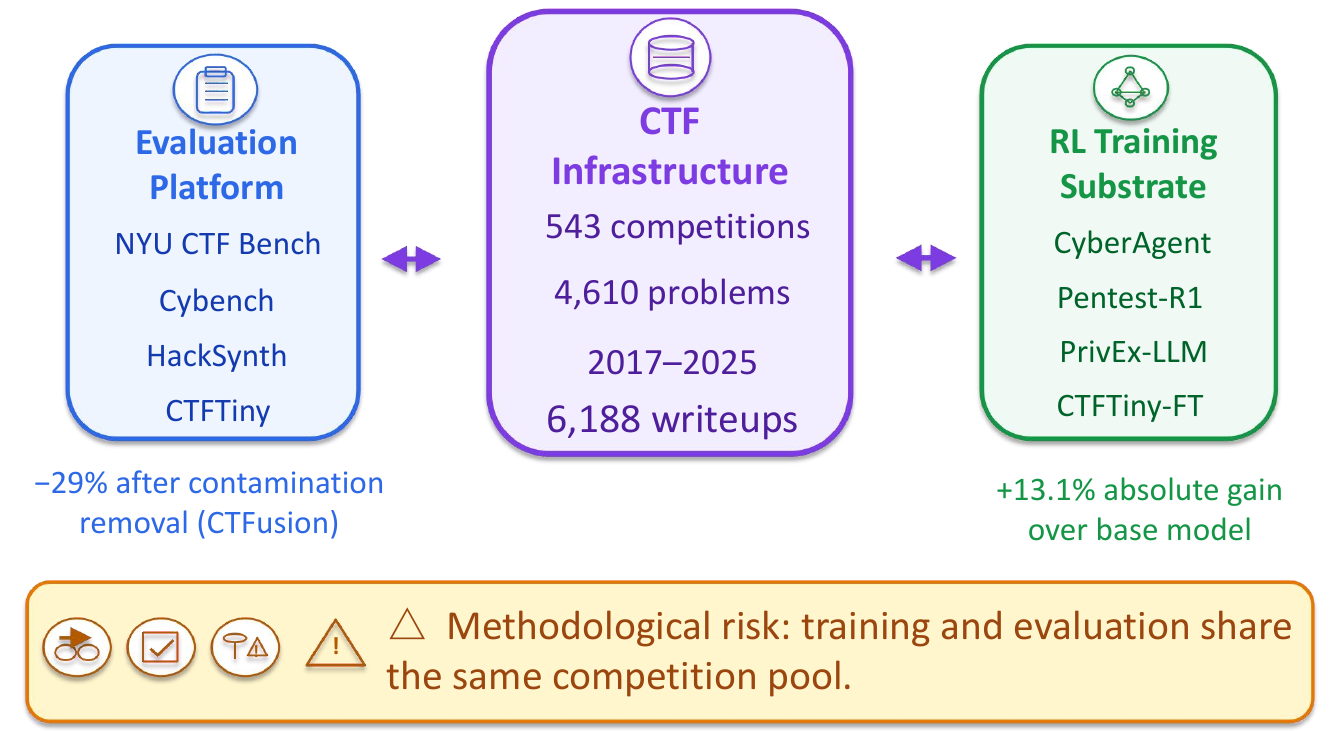}
\caption{CTF infrastructure plays a dual role in Agent4Pentest research pipeline, serving as both an evaluation platform and a RL training substrate. The same competition challenges, writeup archives, and containerized execution environments support both uses. This tight coupling introduces a methodological risk: agents fine-tuned on CTFtime writeups and evaluated on challenges from overlapping competition events may exhibit benchmark gains that reflect format familiarity rather than genuine generalization.}
\label{fig:ctf-dual}
\end{figure}

After each competition, participants often publish writeups that document their solution processes, including the identified vulnerability, tools used, dead ends encountered, and final steps that produced the flag. This practice repurposes CTF archives from evaluation material into training data, leveraging a large collection of high-quality, reasoning-rich solution records accumulated across hundreds of public competitions and diverse vulnerability types. 
Fig.~\ref{fig:ctf-dual} illustrates this dual role of CTF infrastructure and highlights the methodological risk that arises when the training corpus and evaluation tasks are drawn from overlapping competition pools.

One work\cite{zhuo2025cyber} exploits this property through dual-model simulation, in which a competitor model proposes investigation steps and a terminal model synthesizes plausible system responses, producing trajectories with trial-and-error cycles and strategy revisions without requiring a real execution environment. The training corpus contains 6,188 high-quality writeups from CTFtime~\cite{krause2008musculoskeletal}, covering 543 competitions and 4,610 unique problems from 2017 to 2025. A Qwen3-32B model fine-tuned on this data achieves a 13.1\% absolute improvement over the base model across three standard CTF benchmarks, reaching performance comparable to \mbox{Claude-3.5-Sonnet}\cite{templeton2026scaling} and \mbox{DeepSeek-V3-0324}\cite{liu2024deepseek} at substantially lower inference cost. 
Prior work on capable CTF agents relied primarily on prompting large proprietary models or collecting manually labeled trajectories through costly human effort; this study\rev{\cite{zhuo2025cyber}} shows that public writeup archives can partially substitute for both. However, the approach also inherits distributional bias from the writeup corpus, where successful attack paths are overrepresented relative to failed attempts. Generating counterfactual failure traces to correct this bias remains an open problem. 

More broadly, CTF-based Agent4Pentest systems show that CTF infrastructure supports the research pipeline beyond evaluation alone: the same challenge sets, writeup archives, and containerized environments can support both benchmark evaluation and agent training. This shared infrastructure creates an unusually tight coupling between training and evaluation compared with other areas of AI security research. On productive side, improvements in challenge diversity and writeup quality directly expand the available training distribution. On methodological side, agents trained and evaluated on challenges from overlapping competition events may show benchmark improvements that reflect format familiarity rather than generalization to novel scenarios. Thus, the field needs standard protocols for separating training and evaluation task sets, especially as trained CTF agents become more capable.
\section{General-purpose Agent4Pentest Systems}\label{sec:autopt}

This section surveys the general-purpose Agent4Pentest systems in our Category~II.
We organize the analysis along three dimensions: architectural evolution (\S\ref{subsec:autopt-evolution}), cross-cutting system components (\S\ref{subsec:autopt-components}), and training paradigms (\S\ref{subsec:autopt-training}).

\subsection{Architecture Overview}\label{subsec:autopt-evolution}

By analyzing the 36 general-purpose Agent4Pentest systems chronologically, we organize Category~II systems into four architectural types: text-only reasoning agents, tool-augmented single agents, multi-agent coordination systems, and RLVR-trained agents. These types correspond to the four-phase progression (\S\ref{subsec:evolution}). Table~\ref{tab:autopt-systems} summarizes representative systems, their architectures, key innovations, and best reported results. We next discuss each architectural type in turn.


\noindent\textbf{(i) Text-only reasoning agents} represent the earliest design of  Agent4Pentest, demonstrating that structured state representation can substantially improve attack reasoning over direct LLM prompting. Their defining feature is a task tree that organizes an engagement into navigable attack branches, enabling the agent to track open objectives and accumulate findings across turns. However, these systems still rely on a human operator to execute every proposed command and relay results back to the model. PentestGPT~\cite{deng2024pentestgpt} exemplifies this design with a three-module architecture: a Reasoning Module maintains a Pentesting Task Tree (PTT) to track engagement state, a Generation Module converts each PTT node into concrete action steps, and a Parsing Module extracts relevant findings from raw tool output relayed by the human operator. Evaluated on 13 HackTheBox~\cite{hackthebox26} and VulnHub targets~\cite{vulnhub26}, PentestGPT improves subtask completion by 58.6\% over direct GPT-4 use, establishing structured state management as a durable design principle. The unresolved bottleneck is \textit{execution autonomy}: every proposed command still depends on human execution and transcription.

\noindent\textbf{(ii) Tool-augmented single agents} address execution autonomy by equipping a single LLM direct access to security tools, thereby eliminating the human relay required by text-only systems. These architectures typically combine high-level planning, direct tool invocation, and knowledge retrieval from vulnerability databases to configure exploits with minimal manual input. PentestAgent~\cite{shen2025pentestagent} exemplifies this design with a four-agent pipeline that integrates online CVE retrieval, enabling the agent to automatically configure public exploits for discovered services. It achieves 74.2\% overall success across a 67-target benchmark. The broader set of tool-augmented single-agent systems~\cite{nakatani2025rapidpen,huang2023penheal,ginige2025autopentester,fang2024llm,bianou2024pentest,al2025pentest++,xu2024autoattacker,skandylas2025automated} shows meaningful capability gains on short engagements, but performance degrades as engagement length grows and context accumulates. The unresolved bottleneck therefore shifts to \textit{context management}.

\noindent\textbf{(iii) Multi-agent coordination systems} address context degradation by distributing the engagement across role-specialized subagents, each maintaining a focused context window for its assigned phase. Their defining feature is a task or attack-graph representation that encodes dependencies among subtasks, typically combined with cross-agent reflection or recovery mechanisms that enable targeted failure correction without restarting the full engagement. Incalmo targets multi-host enterprise red teaming by decoupling an LLM planning layer from expert execution agents and using an attack-graph service to maintain a persistent asset register as the agent moves laterally across host boundaries~\cite{singer2026incalmo}. Evaluated on 40 simulated enterprise-network scenarios in MHBench, Incalmo succeeds on 37 environments, whereas the strongest single-agent baseline succeeds on only 3. Ablation studies further show that planning–execution decoupling contributes more to performance than model scale. The wider multi-agent literature~\cite{wang2025automated,alshehri2024breachseek,geng2025controller,david2025multi,mai2025shell,khang2026red,dai2025refpentester,wang2025ptfusion,zhai2025pentestmcp,jaswalawe2026awe,luong2025xoffense,deng2026makes,mayoral2025cai,challita2025redteamllm} similarly shows that focused subtask contexts reduce context overload and per-step failure rates across graph-based, state-machine-based, and orchestrator–worker implementations. The remaining bottleneck is \textit{training-data scarcity}: most agents still learn from human demonstrations, constraining the learnable strategy space to techniques already recorded by human testers.

\noindent\textbf{(iv) RLVR-trained agents} address training-data scarcity by replacing purely demonstration-based learning with verifiable environment rewards. Instead of relying only on curated trajectories, these agents explore live targets and receive reward signals derived from outcomes such as successful exploitation. Their training typically combines supervised fine-tuning to seed basic attack knowledge with online reinforcement learning, allowing the agent to discover strategies beyond the human-curated corpus. Pentest-R1~\cite{kong2025pentest} applies a two-stage framework: an offline stage fine-tunes the model on 500 expert walkthroughs, and an online Group Relative Policy Optimization (GRPO)~\cite{shao2024deepseekmath} stage trains the agent in interactive CTF environments to develop adaptive error-correction strategies absent from the offline corpus. On AutoPenBench, Pentest-R1 achieves 24.2\% task completion, surpassing GPT-4o and most frontier models; on Cybench, it reaches 15.0\% on unguided tasks, matching top closed-source systems at lower inference cost. The remaining bottleneck is \textit{sample efficiency}: reliable exploitation requires many attack attempts, and resetting the target environment after each attempt is costly at scale.

\begin{figure}[t]
\centering
\includegraphics[width=\linewidth]{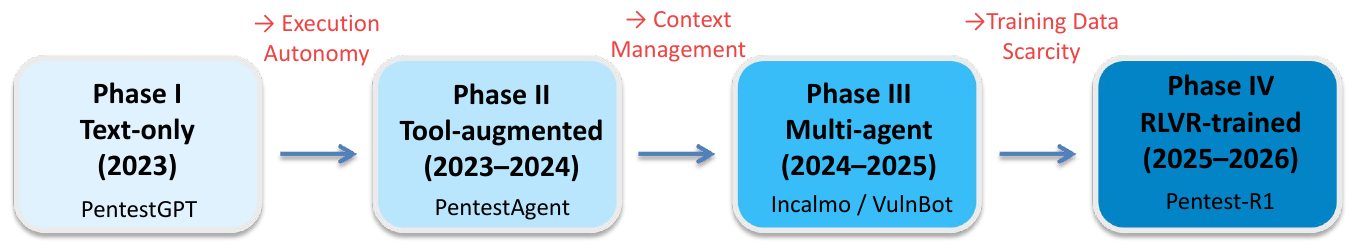}
\caption{General-purpose Agent4Pentest systems' four-phase architectural progression.}
\label{fig:autopt-phases-timeline}
\end{figure}

\begin{figure}[t]
\centering
\includegraphics[width=\linewidth]{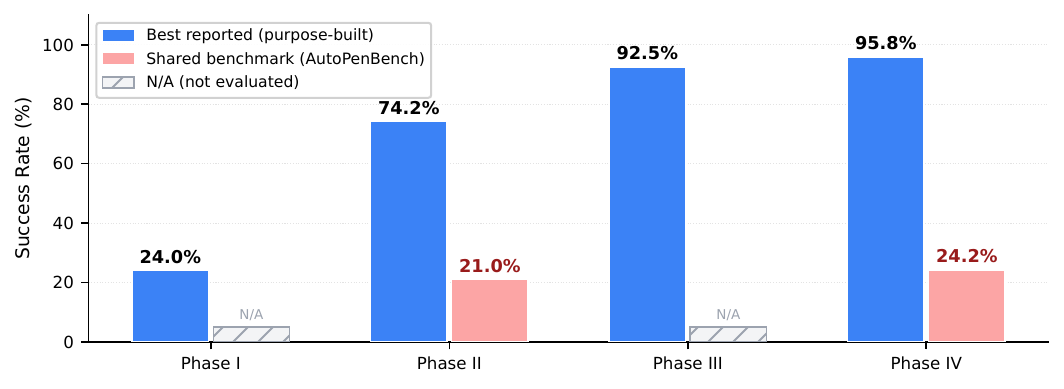}
\caption{The rates of task completion by architectural phase on purpose-built benchmarks (blue bars) and, where available, the shared AutoPenBench benchmark~\cite{gioacchini2025autopenbench} (red bars). PhaseI reports 24.0\% on HackTheBox\cite{hackthebox26} and VulnHub~\cite{vulnhub26} targets, corresponding to a 58.6\% relative gain over GPT-4. PhaseII reaches 74.2\% on a 67-target purpose-built suite\cite{shen2025pentestagent} and 21.0\% on AutoPenBench. PhaseIII achieves 92.5\% on MHBench\cite{singer2026incalmo}, solving 37 of 40 tasks. PhaseIV reaches 95.8\% on a Linux privilege-escalation suite\cite{normann2026post} and 24.2\% on AutoPenBench. Purpose-built and shared benchmarks are not directly comparable; they are shown together only to illustrate within-phase performance scales.}
\label{fig:autopt-phases}
\end{figure}

In summary, each architectural type resolves one bottleneck while exposing the next, indicating that capability advances arise from specific architectural improvements rather than model scaling alone. This progression has driven rising task-completion rates, with multi-agent and RLVR-trained systems achieving scores that earlier single-agent designs could not reach on shared benchmarks. Reported success rates, however, vary substantially across evaluation settings: some systems exceed 70\% on purpose-built benchmarks but fall below 30\% on shared benchmarks such as AutoPenBench, suggesting that purpose-built evaluations may overestimate real-world capability. Fig.~\ref{fig:autopt-phases-timeline} summarizes the four-phase architectural progression and the bottleneck each phase resolves and exposes. Fig.~\ref{fig:autopt-phases} shows representative task-completion rates per phase on purpose-built benchmarks and, where available, on the shared AutoPenBench~\cite{gioacchini2025autopenbench}.

\begin{table}[t!]
\centering
\caption{Representative general-purpose Agent4Pentest systems from each architectural phase, selected for individual discussion in \S\ref{subsec:autopt-evolution}.
Remaining papers in each phase are cited as a group in the text.}
\label{tab:autopt-systems}
\begingroup
\renewcommand{\arraystretch}{1.15}
\resizebox{\linewidth}{!}{%
\begin{tabular}{@{}lllll@{}}
\hline
\textbf{Phase} & \textbf{System} & \textbf{Architecture} & \textbf{Key Innovation} & \textbf{Best Result} \\
\hline
I  & PentestGPT~\cite{deng2024pentestgpt}
   & 3-module, text-only
   & Pentesting Task Tree (PTT)
   & $+$58.6\% vs.\ direct GPT-4 \\
II & PentestAgent~\cite{shen2025pentestagent}
   & 4-agent pipeline
   & Online CVE retrieval
   & 74.2\% (67 targets) \\
III & Incalmo~\cite{singer2026incalmo}
   & Plan-exec decoupled
   & Attack-graph service
   & 37/40 (MHBench) \\
IV & Pentest-R1~\cite{kong2025pentest}
   & 2-stage GRPO
   & Verifiable-reward RL
   & 24.2\% (AutoPenBench) \\
\hline
\end{tabular}}
\endgroup
\end{table}

\subsection{Core Component Analysis}\label{subsec:autopt-components}

\begin{figure}[t]
\centering
\includegraphics[width=\linewidth]{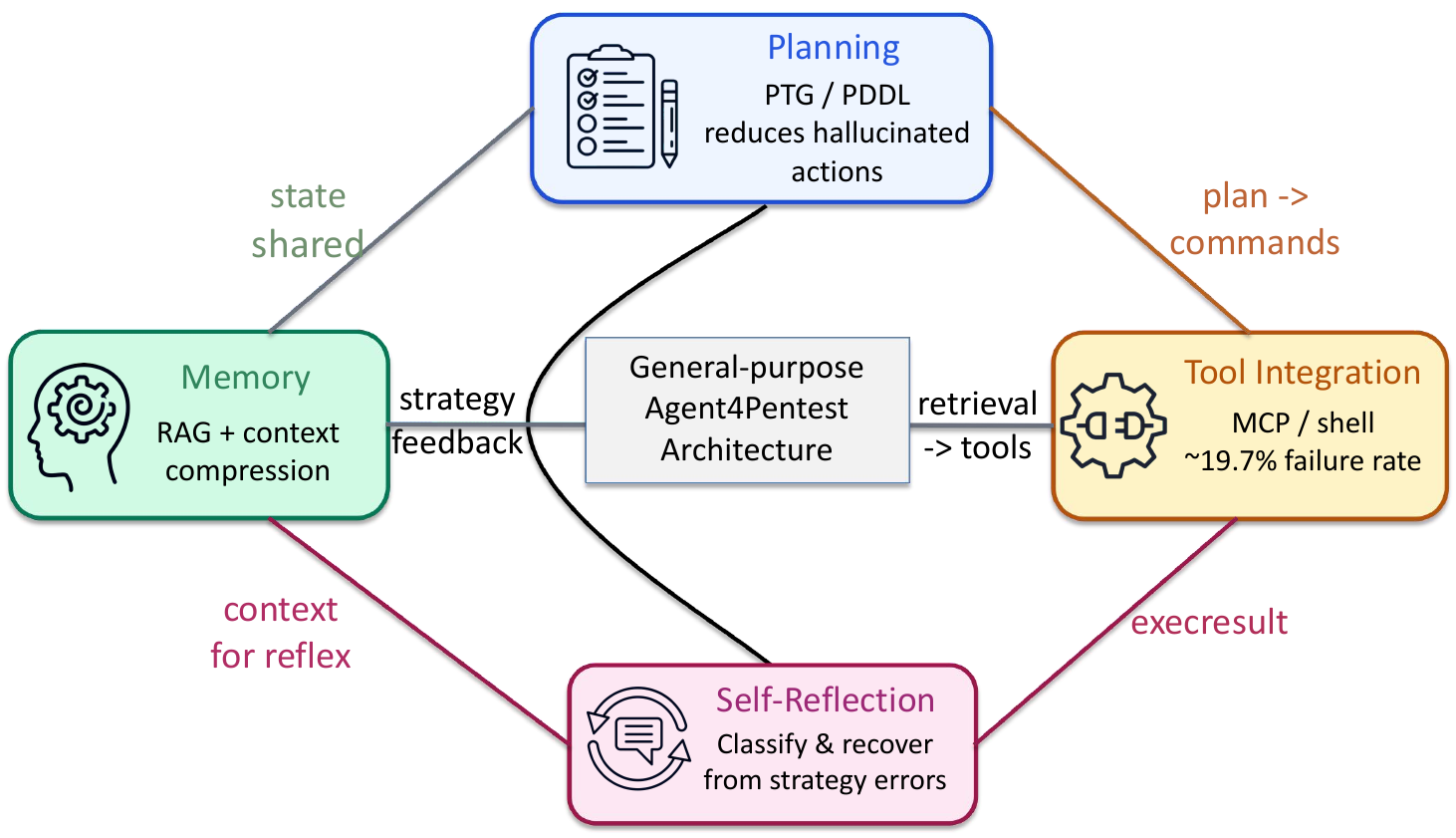}
\caption{Four core components of general-purpose Agent4Pentest systems and their interdependencies. Planning reduces hallucinated actions by maintaining structured attack state; memory prevents context degradation and enables domain knowledge retrieval; tool integration reduces execution failures, which account for approximately 19.7\% of all failures in published ablations~\cite{kong2025vulnbot}; and self-reflection recovers from strategy errors without restarting the engagement. Each component addresses a distinct failure mode that the others cannot compensate for.}
\label{fig:autopt-components}
\end{figure}

We further identify four core components that recur across the 36 general-purpose Agent4Pentest systems: (i) a planning mechanism for high-level attack strategy, (ii) a memory system for context management and domain knowledge retrieval, (iii) a tool-integration layer that connects decisions to executable commands, and (iv) a self-reflection mechanism for failure recovery. Fig.~\ref{fig:autopt-components} shows the dependencies among these components, the data flows between them, and the distinct failure mode addressed by each component.

The planning mechanism maintains a structured representation of the current attack state and decomposes the overall objective into executable subtasks, guiding decisions across the engagement. Planning representations have evolved from task trees to graph-based structures and externally verified planners. The Pentesting Task Tree (PTT) introduced by PentestGPT~\cite{deng2024pentestgpt} serializes attack strategy as a navigable tree, but its topology makes it difficult to represent parallel subtasks or loops back to earlier nodes. 
The Pentesting Task Graph (PTG) adopted by VulnBot~\cite{kong2025vulnbot} and related systems addresses this limitation with a directed acyclic graph, allowing the planner to schedule independent subtasks in parallel and encode conditional dependencies between vulnerability discovery and exploitation. 
CHECKMATE~\cite{wang2025automated} takes a different route by replacing the neural planner with a classical PDDL-based planner~\cite{zhang2025lamma}, augmented by an LLM that translates scan outputs into symbolic predicates. 
Nakano et al.\cite{nakanoguided} constrain the LLM to select from a predefined MITRE ATT\&CK task tree\cite{xiong2022cyber}, demonstrating that external taxonomic structure reduces hallucinated actions and improves subtask completion. 
Overall, explicit planning structure consistently reduces hallucinated actions and improves task completion compared with unguided LLM prompting, suggesting that planning is better handled through structured representations than through free-form reasoning alone.

The memory system stores and retrieves information needed across an engagement. It serves two distinct functions that require separate mechanisms: context management and domain knowledge retrieval. Context management addresses the problem that long engagements fill the active context window with irrelevant tool output and degrade reasoning quality. Systems such as VulnBot~\cite{kong2025vulnbot} and PenHeal~\cite{huang2023penheal} address this issue by inserting a Summarizer that compresses each tool response into a compact finding before appending it to the context. Domain knowledge retrieval addresses the need to identify the correct exploitation procedure for a given service or vulnerability class during the engagement. RAG-based modules are used across multiple systems~\cite{shen2025pentestagent,ginige2025autopentester,dai2025refpentester} to retrieve relevant entries from static knowledge bases encoding OWASP guidelines~\cite{lala2021secure}, HackTricks procedures~\cite{carlos2024hacktricks}, and CVE exploit details. A more dynamic variant maintains the knowledge base as a live graph updated as new hosts and services are discovered~\cite{wang2025ptfusion,li2026intelligent}, allowing the planner to query the evolving attack surface rather than a fixed index. Context compression and knowledge retrieval pull in opposite directions; systems that conflate them tend to underperform those that maintain separate mechanisms for each.

The tool-integration layer connects agent decisions to executable security tools such as Nmap~\cite{orebaugh2011nmap}, Metasploit~\cite{kennedy2011metasploit}, and SQLMap~\cite{axinte2014sql}. Its reliability directly determines whether the agent can act on its plan. Early tool-augmented single agents invoke security tools by constructing shell commands in LLM outputs and parsing the resulting text, a fragile approach in which minor changes in tool-output formatting can break the parsing pipeline. Multi-agent coordination systems increasingly adopt the Model Context Protocol (MCP)\cite{hou2026model,hasan2025model} to expose security tools as structured function calls\cite{wang2025ptfusion,zhai2025pentestmcp}, reducing integration failures. TermiAgent~\cite{mai2025shell} addresses tool coverage by packaging 1,378 CVE-specific exploits as uniformly invocable Docker containers, expanding coverage to 694 RCE CVEs compared with the 468 provided by Metasploit alone. We observe that approximately 19.70\% of engagement failures~\cite{kong2025vulnbot} stem from tool-invocation errors rather than reasoning failures, establishing tool reliability as an open engineering challenge independent of model capability.

The self-reflection mechanism allows the agent to assess whether an action succeeded and adjust its strategy accordingly, preventing repeated execution of the same failing command. Text-only reasoning agents lack failure recovery and generate the next step regardless of whether the previous action produced a useful result. Tool-augmented single agents add retry logic that repeats the same command with a modified prompt, but they often do so without diagnosing the root cause of failure. The Check-and-Reflection mechanism in VulnBot~\cite{kong2025vulnbot} improves on this design by classifying each failure as either a tool error or a strategy error and dispatching a targeted recovery procedure instead of applying a uniform retry. Red-MIRROR~\cite{khang2026red} extends reflection to two levels, varying payload encodings within a single turn while applying majority-vote verification across turns before advancing. Consistent performance gains at each level of sophistication indicate that failure diagnosis is as important as planning for long-horizon attack tasks, and that agents without reflection waste substantial budget repeating failed actions.

We draw two observations from this component-level analysis. First, none of the four components can be safely omitted, because each addresses a distinct failure mode that the others cannot compensate for. Second, these components are interdependent: a more expressive planning structure requires a stronger memory system to track its state, while a more reliable tool layer reduces the burden on reflection to recover from execution errors.

\subsection{Training Paradigms}\label{subsec:autopt-training}

Training paradigms used in general-purpose Agent4Pentest systems differ primarily in the signal that drives capability acquisition. Text-only reasoning agents and tool-augmented single agents rely mainly on prompt engineering: the backbone model is selected from a commercial or open-source provider, and behavior is shaped through task-specific prompts without modifying model weights. Multi-agent coordination systems introduce supervised fine-tuning (SFT) on penetration-testing walkthroughs and domain-specific corpora, enabling smaller models to acquire specialized attack knowledge not reliably captured by generic pretraining. RLVR-trained agents further shift the training signal from human-labeled demonstrations to verifiable environment rewards, allowing agents to discover attack strategies beyond the human-curated corpus.

A consistent finding across this literature is that small specialist models fine-tuned on penetration-testing data can match or exceed much larger general-purpose models at substantially lower inference cost. A 7B specialist model fine-tuned on more than 300 HackTheBox writeups and security-tool documentation~\cite{pratama2024cipher} outperforms Llama-3-70B on the FARR Flow penetration-testing reasoning benchmark. LoRA fine-tuning of Qwen3-32B on 1,000 penetration-testing walkthroughs~\cite{luong2025xoffense} raises AutoPenBench task completion to 72.72\%, more than doubling the performance of the same base model without fine-tuning. Applying GRPO to a Qwen-3-14B strategy model~\cite{ginige2026pen} increases its GEval strategy score from 0.39 to 0.73 and improves subtask completion across three integrated frameworks by 47.5\% on average. The same two-stage SFT-then-GRPO approach applied to translating natural language into Kali Linux commands~\cite{kong2026intent} achieves higher exact-match accuracy than models with an order of magnitude more parameters.

\begin{figure}[t]
\centering
\includegraphics[width=\linewidth]{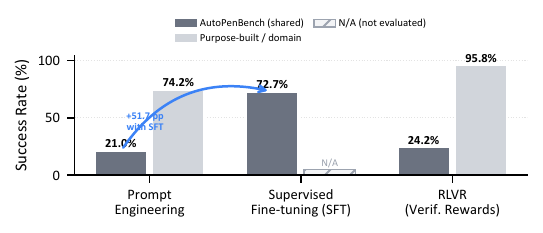}
\caption{Performance of representative Agent4Pentest systems across three training paradigms on AutoPenBench (shared benchmark, dark bars) and on the best purpose-built or domain benchmark where available (light bars). Each bar represents a different system: PentestAgent~\cite{shen2025pentestagent} (GPT-4o, prompt engineering), Qwen3-32B LoRA~\cite{luong2025xoffense} (SFT on 1,000 domain walkthroughs), and Pentest-R1~\cite{kong2025pentest} (GRPO-based RLVR). The 95.8\% domain result for RLVR comes from PrivEx-LLM~\cite{normann2026post}, a separate RLVR system targeting Linux privilege escalation rather than Pentest-R1.}
\label{fig:autopt-training}
\end{figure}

RLVR produces the strongest capability gains by training directly on verifiable environment signals rather than relying on human-labeled demonstrations as the primary learning source. Pentest-R1~\cite{kong2025pentest} shows that online GRPO is essential: removing this stage and retaining only offline SFT substantially degrades performance, confirming that environment rewards drive agents to discover exploit sequences absent from human demonstrations. A follow-on study on Linux privilege escalation~\cite{normann2026post} designs five reward components that encode terminal success, interaction speed, reconnaissance quality, and penalties for invalid or repeated commands. This reward design trains a 4B model to reach 95.8\% success within 20 interaction rounds, closely matching Opus~\cite{choi2025comparison} at more than 100 times lower inference cost. Together, these results suggest that reward design, rather than model scale or demonstration provenance alone, is the key driver of RLVR capability gains.
Fig.~\ref{fig:autopt-training} shows representative systems from three training paradigms on AutoPenBench (dark bars) and on the best purpose-built or domain benchmark where available (light bars). AutoPenBench results should not be read as a clean paradigm progression, as the compared systems use different training distributions and target different task settings.

In summary, general-purpose Agent4Pentest systems have progressed from text-only advisory tools to reward-trained autonomous operators in less than four years. Across this progression, architectural consensus has emerged around structured planning, RAG-enhanced memory, standardized tool integration, and reflection-driven error recovery. The main open challenge for training is sample efficiency in RLVR: successful attack episodes are sparse in realistic environments, limiting the discovery of novel strategies without either large episode budgets or more efficient exploration methods.

\section{Domain-specific Agent4Pentest Frameworks}\label{sec:domain}

This section surveys the domain-specific Agent4Pentest frameworks in Category~III of our taxonomy. We organize the analysis along three dimensions: domain coverage, which maps systems to attack scenarios (\S\ref{subsec:domain-coverage}); architectural specialization, which identifies mechanisms shared across systems (\S\ref{subsec:domain-arch}); and specialization limits, which examines the trade-offs introduced by narrowing system scope (\S\ref{subsec:domain-limits}).

\begin{table}[b!]
\centering
\caption{Representative domain-specific Agent4Pentest frameworks}
\label{tab:domain-systems}
\begingroup
\footnotesize
\setlength{\tabcolsep}{1.5pt}
\renewcommand{\arraystretch}{1.25}
\begin{tabular}{@{}C{0.25\linewidth}C{0.24\linewidth}C{0.25\linewidth}C{0.18\linewidth}@{}}
\hline
\textbf{Domain / System} &
\textbf{Architecture} &
\textbf{Key Innovation} &
\textbf{Best Result} \\
\hline
Privilege Escalation\newline ChainReactor~\cite{de2024chainreactor}
  & PDDL + classical\newline planner
  & Automated chain\newline discovery
  & 16 chains on\newline real EC2/DO\newline instances \\
Privilege Escalation\newline PrivEx-LLM~\cite{normann2026post}
  & SFT + RLVR,\newline two-stage
  & Verifiable-reward\newline fine-tuning
  & 95.8\% (Linux,\newline 20 rounds) \\
Enterprise Network\newline Happe et al.~\cite{happe2026can}
  & Multi-phase,\newline MITRE ATT\&CK
  & Assumed-breach\newline AD testing
  & Cost competitive\newline with human\newline testers \\
Web Application\newline AWE~\cite{jaswalawe2026awe}
  & 3-layer orchestration
  & Vuln-class\newline specialist agents
  & 87\% XSS,\newline 66.7\% blind SQLi \\
\hline
\end{tabular}
\endgroup
\end{table}

Inspired by the success of domain-specific language models~\cite{cook2007domain}, we define this category as follows: a \textit{domain-specific Agent4Pentest} framework is a system that restricts its scope to a single attack scenario or vulnerability class and is designed and evaluated within that domain rather than across multiple engagement types or general-purpose benchmarks. This restriction allows each system to replace generic LLM reasoning with toolchains, knowledge bases, and attack workflows tailored to the target domain. Table~\ref{tab:domain-systems} lists the representative systems selected for detailed discussion, while the remaining papers in each domain are cited collectively.

\subsection{Domain Coverage}\label{subsec:domain-coverage}

\begin{figure}[t]
\centering
\includegraphics[width=\linewidth]{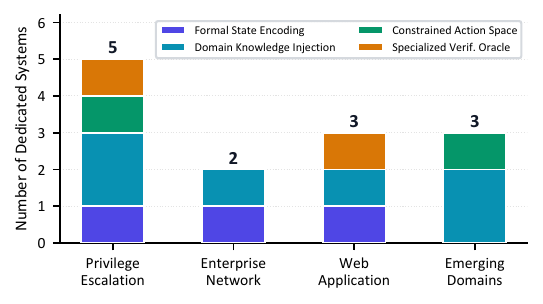}
\caption{Attack-domain coverage and primary specialization mechanisms across the 13 domain-specific Agent4Pentest frameworks. Bar height indicates the number of dedicated systems per domain, and colored segments indicate the primary architectural specialization mechanism used by each system. 
}
\label{fig:domain-coverage}
\end{figure}

We first examine the domain coverage of existing domain-specific Agent4Pentest frameworks, including privilege escalation, enterprise network compromise, web application penetration testing, and emerging domains with only one dedicated system. We summarize these domains below. Fig.~\ref{fig:domain-coverage} summarizes the domain distribution and architectural mechanisms used across the 13 papers in Category~III.

\noindent\textbf{Privilege Escalation} aims to obtain higher system privileges from an initial low-privilege shell by exploiting misconfigurations, outdated software, or insecure file permissions on the target host. ChainReactor~\cite{de2024chainreactor} follows a perceive-plan-act architecture that encodes target system state as PDDL facts and uses a classical AI planner to generate multi-step exploitation chains. On real Amazon EC2 and DigitalOcean instances, it rediscovers 16 known privilege-escalation chains and identifies previously unreported ones. Its agent loop is architecturally analogous to LLM-based Agent4Pentest systems, and we include it as the primary non-LLM baseline for evaluating LLM-based privilege-escalation agents. PrivEx-LLM~\cite{normann2026post} applies two-stage RLVR training with five reward components encoding terminal success, interaction speed, and reconnaissance quality, reaching 95.8\% success on Linux privilege escalation at more than 100 times lower inference cost than Claude Opus. Perses~\cite{weber2025perses} takes a complementary approach by combining heterogeneous small models through role-based assignment, achieving 87.5\% success on FreeBSD privilege escalation without using any frontier model.

\noindent\textbf{Enterprise Network Compromise} targets the authentication and authorization infrastructure of corporate networks, typically Active Directory, with the goal of obtaining domain-administrator access through multi-hop credential theft and lateral movement. 
The framework of Happe et al.\cite{happe2026can} is the first fully autonomous LLM-driven system for assumed-breach penetration testing in real enterprise Active Directory environments. It organizes the engagement into reconnaissance, credential access, and lateral movement phases following MITRE ATT\&CK, and finds that deployment costs are competitive with professional human testers while raising significant safety concerns about autonomous offensive agents in live environments. 
CHIMERA\cite{yu2026chimera} addresses the complementary problem of generating realistic insider-threat behavior for evaluating detection systems. It models each employee as an LLM-powered agent under realistic role-based access controls and generates a synthetic dataset of 25 billion log entries, revealing that existing insider-threat detection models generalize poorly under changes in organizational context.

\noindent\textbf{Web Application Penetration Testing}
targets injection-class vulnerabilities in web services, including server-side template injection~\cite{bhuiyan2023secbench}, cross-site scripting~\cite{grossman2007xss}, and SQL injection~\cite{halfond2006classification}. Each vulnerability class requires a distinct payload structure and domain-specific verification method. In services computing areas, these vulnerabilities are especially dangerous because a successful injection at a single REST or GraphQL API endpoint can propagate across service-composition boundaries and reach protected backend resources~\cite{lin2010dynamic,murakami2012service}. AWE~\cite{jaswalawe2026awe} organizes the agent into three layers: an Orchestration Layer for global state management, a Specialized Agents Layer with one agent per vulnerability class, and a Foundation Layer that provides persistent memory and browser-backed verification. On XBOW benchmark, AWE achieves 87\% success on XSS and 66.7\% on blind SQL injection~\cite{sharma2014analysis}, improving over the strongest multi-agent baseline by 30.5\% and 33.3\%, respectively, while reducing token consumption by 98\%. The broader set of web-focused systems~\cite{david2025multi,shao2026towards} confirms that combining structured vulnerability knowledge with LLM orchestration yields consistent gains over general-purpose agents on narrow task categories.

\noindent\textbf{Emerging Attack Domains}, including SSH-shell penetration testing~\cite{barrett2005ssh}, wireless network assessment, and embedded systems security, are each addressed by only one dedicated system in CategoryIII, indicating that they remain at an early stage of research attention. ARACNE\cite{nieponice2025aracne} targets SSH-based Linux compromise using a multi-LLM architecture that separates planning from command generation, achieving 60\% success against a live honeypot. WiFiPenTester~\cite{al2026wifipentester} addresses IEEE 802.11~\cite{hameed2014lte} wireless assessment under a governed-autonomy model that requires operator approval before command execution, preserving legal accountability while benefiting from LLM-assisted reconnaissance. A benchmarking framework for ARM-based edge and IoT systems~\cite{ragsdale2026ai,kaiser2023benchmarking,kayan2025real} finds that LLM configurations exceed practical memory and latency budgets for edge deployment, whereas reinforcement learning combined with model quantization provides a viable lightweight alternative~\cite{nan2017adaptive}.

We draw two observations from this domain-coverage survey. First, the distribution is uneven: privilege escalation attracts the most dedicated systems because its attack state is fully observable and its success condition is precisely defined, whereas enterprise network compromise and web testing attract fewer systems because their attack surfaces are open-ended and harder to formalize. Second, dedicated frameworks tend to emerge when an attack workflow becomes sufficiently stereotyped to be encoded as a fixed knowledge base or reward function. Privilege escalation crossed this threshold first, whereas SSH testing, wireless assessment, and embedded security have not yet done so. As more attack workflows are formalized through benchmark development and CTF-style environment construction, dedicated domain-specific frameworks are likely to emerge for these currently underserved scenarios.

\subsection{Architectural Specialization Mechanisms}\label{subsec:domain-arch}

\begin{figure}[t]
\centering
\includegraphics[width=\linewidth]{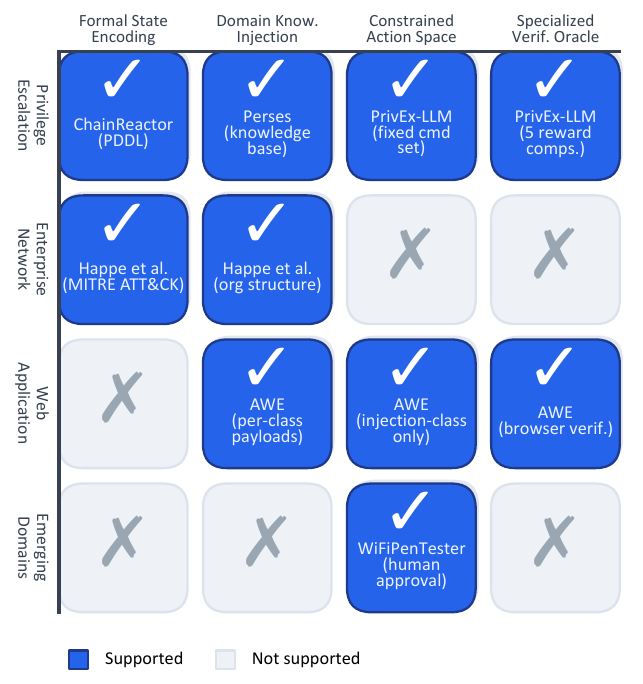}
\caption{Presence of four architectural specialization mechanisms across the four attack domains. A filled cell indicates that at least one system in that domain employs the mechanism, with the representative system annotated inside the cell. Privilege escalation is the only domain that uses all four mechanisms, reflecting its fully observable attack state and precisely defined success criterion.}
\label{fig:domain-mechanisms}
\end{figure}

We find that the architectural choices of domain-specific Agent4Pentest frameworks fall into four recurring specialization mechanisms: formal state encoding, domain knowledge injection, constrained action spaces, and specialized verification oracles. These mechanisms appear across all four attack domains despite differences in target systems and vulnerability types. Fig.~\ref{fig:domain-mechanisms} maps the co-occurrence of these mechanisms across attack domains, showing that formal state encoding and verification oracles cluster in domains where the attack state is fully observable and the success condition is precisely defined.

\noindent\textbf{Formal state encoding.}
The most distinctive specialization mechanism replaces open-ended LLM reasoning with a formal representation of the attack state. ChainReactor~\cite{de2024chainreactor} encodes system configurations as PDDL facts and delegates chain discovery to a classical planner, replacing the neural reasoning component with a symbolic one while preserving the agent’s perceive-plan-act structure. Happe et al.~\cite{happe2026can} constrain the agent to reason within the MITRE ATT\&CK taxonomy, reducing the action space from arbitrary shell commands to a structured set of technique identifiers. We observe that formal state encoding is most effective in domains where the attack state is fully observable and the goal condition is precisely defined, such as privilege escalation and Active Directory lateral movement.

\noindent\textbf{Domain knowledge injection.}
Systems across all four domains augment the LLM with static or dynamic knowledge bases that encode domain-specific exploitation procedures. For privilege escalation, Perses~\cite{weber2025perses} integrates a knowledge base of known misconfigurations and exploit procedures indexed by system binary. For web testing, AWE~\cite{jaswalawe2026awe} embeds per-vulnerability-class payload mutation strategies directly into specialized agents. For enterprise compromise, Happe et al.~\cite{happe2026can} populate the agent with organizational structure inferred during reconnaissance, enabling targeted credential reuse across hosts. We observe that static knowledge bases are sufficient for closed vulnerability classes such as privilege-escalation misconfigurations, whereas dynamic knowledge graphs are necessary for open-ended domains such as enterprise lateral movement, where the attack surface changes as the agent progresses.

\noindent\textbf{Constrained action spaces.}
Domain-specific systems restrict the commands an agent can issue to those relevant to the target domain, reducing hallucinated or irrelevant actions. PrivEx-LLM~\cite{normann2026post} limits the agent to a fixed set of privilege-escalation enumeration and exploitation commands, and penalizes repeated or invalid commands through the reward function. WiFiPenTester~\cite{al2026wifipentester} constrains the action space at the system level by requiring human approval before command execution, effectively limiting the blast radius of incorrect actions. We observe that action-space constraints reduce wasted token budget on irrelevant exploration and constitute a major source of the efficiency gains that domain-specific systems achieve over general-purpose baselines.

\noindent\textbf{Specialized verification oracles.}
Domain-specific systems replace the binary flag-capture signals common in general-purpose evaluation with richer, domain-specific verification mechanisms. AWE~\cite{jaswalawe2026awe} uses a live browser to verify that each payload actually executes in the target application, eliminating false positives from LLM-generated exploit code that appears correct but fails at runtime. PrivEx-LLM~\cite{normann2026post} designs five reward components that encode not only terminal success but also intermediate quality signals such as reconnaissance coverage and command efficiency. We observe that richer verification oracles can accelerate RLVR training by providing denser reward signals, and that oracle design is a primary engineering bottleneck for extending RLVR to new domains.

\subsection{Limits of Specialization}\label{subsec:domain-limits}

While specialization consistently improves in-domain task-completion rates, it introduces two structural limitations that distinguish domain-specific frameworks from the general-purpose systems discussed in \S\ref{sec:autopt}: the lack of shared evaluation standards and knowledge brittleness.

\begin{figure}[t]
\centering
\includegraphics[width=\linewidth]{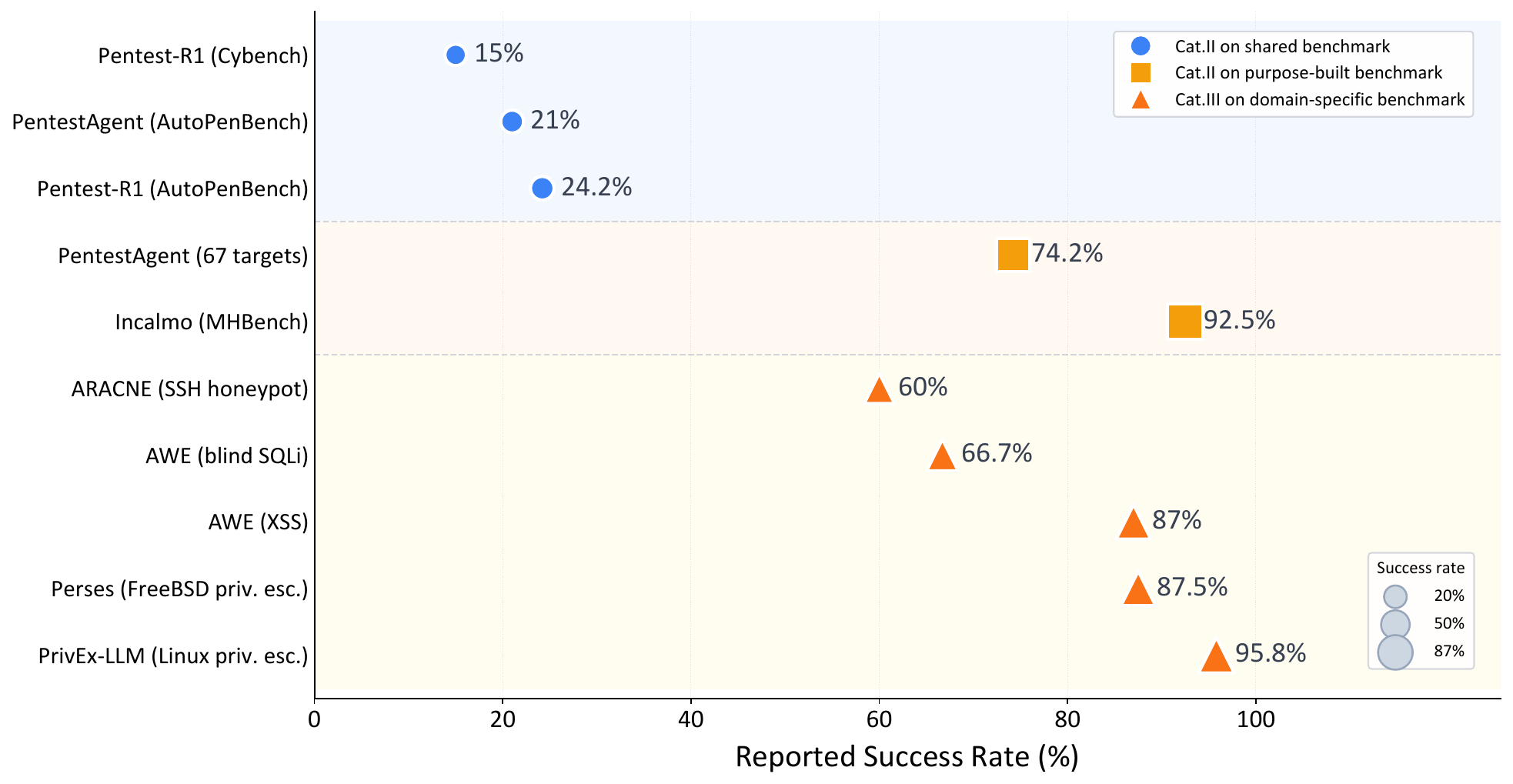}
\caption{Reported success rates across three evaluation settings for Agent4Pentest systems. Left group: general-purpose systems (PentestAgent~\cite{shen2025pentestagent}, Pentest-R1~\cite{kong2025pentest}) on the shared AutoPenBench~\cite{gioacchini2025autopenbench}, reporting 15–24\%. Center group: general-purpose systems (PentestAgent~\cite{shen2025pentestagent}, Incalmo~\cite{singer2026incalmo}) on purpose-built benchmarks, reporting 74–92\%. Right group: domain-specific systems (PrivEx-LLM~\cite{normann2026post}, Perses~\cite{weber2025perses}, AWE~\cite{jaswalawe2026awe}, ARACNE~\cite{nieponice2025aracne}) on their own benchmarks, reporting 60–96\%. The yellow region marks purpose-built evaluations; because the three groups use different benchmarks, direct cross-group capability comparisons are not valid.}
\label{fig:domain-specialization}
\end{figure}

The most direct limitation is evaluation incomparability. Domain-specific systems are typically evaluated on purpose-built benchmarks tailored to their own attack scenarios, whereas general-purpose systems are evaluated on broader shared benchmarks such as AutoPenBench and Cybench. Because the two groups rarely compete on the same tasks, the performance advantages reported by domain-specific systems over general-purpose baselines cannot be interpreted as direct evidence of superior overall capability. Fig.~\ref{fig:domain-specialization} plots reported success rates across three evaluation settings: general-purpose systems on the shared AutoPenBench~\cite{gioacchini2025autopenbench} (15–24\%), general-purpose systems on purpose-built benchmarks (74–92\%), and domain-specific systems on their own benchmarks (60–96\%). PentestAgent~\cite{shen2025pentestagent}, for example, scores 21\% on AutoPenBench but 74.2\% on its own 67-target benchmark, showing a 53.2 percentage-point gap attributable to benchmark choice rather than a change in agent capability.

A second limitation is knowledge brittleness. Domain-specific knowledge bases are effective only within the scope of what they encode: a privilege-escalation knowledge base covering known Linux misconfigurations may not help an agent encountering a novel misconfiguration class, and a web-testing agent trained on injection payloads may not transfer to broken access control or insecure deserialization. General-purpose systems, by contrast, can draw on background knowledge from pretraining to reason about novel vulnerability classes, although usually with lower reliability than specialized systems within their target domains.

Overall, domain-specific frameworks and general-purpose systems occupy complementary positions. Domain-specific systems are most appropriate when the target domain is well defined, the attack surface is bounded, and high task-completion rates on a narrow benchmark are required. General-purpose systems are better suited to engagements whose scope is open-ended or unknown in advance. Thus, the open challenge for domain-specific frameworks is not simply improving in-domain performance, which is already high in several domains, but extending the coverage of their knowledge bases and reward functions to handle the long tail of novel cases outside the training distribution.

\section{Defensive Applications}\label{sec:defense}

This section surveys the papers in Category~V, which represent an emerging research direction in Agent4Pentest systems: adversarial defense and compliance.

The first line of work adopts an adversarial defense approach, disrupting Agent4Pentest systems by exploiting the inherent weaknesses of the attacking model. Cloak Honey Trap~\cite{ayzenshteyn2025cloak} places deceptive prompts and misleading artifacts in the attack surface, causing the attacking agent to follow false leads, waste its token budget, and report incorrect findings to the operator. This strategy works because LLM agents rely heavily on the text they observe; injecting misleading content into the environment can therefore redirect their behavior without modifying the attacker’s model. As offensive agents become more capable, the agent itself becomes a viable attack surface for defenders, since its decisions can be shaped through the documents, prompts, and tool outputs it reads.

The second line of work addresses the accountability gap that arises when an autonomous agent takes offensive actions without continuous human oversight. Intelligent Assurance System~\cite{sanchez2025poster} embeds governance directly into the pentesting loop by monitoring execution traces, auditing each recorded action against EU AI Act~\cite{act2024eu} and GDPR~\cite{sirur2018we} requirements, and injecting compact corrective instructions into the agent’s next planning prompt.
Compliance monitoring becomes more important as agent autonomy increases, because the gap between what an agent is authorized to do and what it is capable of doing widens with each more capable generation of models.

Together, these works show that the Agent4Pentest research community is beginning to examine the defensive and governance implications of its own systems, opening two research directions that prior surveys have not identified. One direction treats deployed offensive agents as adversaries whose decision processes can be disrupted through environmental manipulation. The other treats the pentesting loop itself as a system that requires internal policy enforcement rather than post-hoc human review. Both directions are likely to become more pressing as offensive agents grow more capable and the consequences of uncontrolled autonomous action increase.

\section{Open Challenges and Future Directions}\label{sec:challenges}

Agent4Pentest research has advanced rapidly, but four structural challenges remain unresolved: evaluation reliability, limited performance on multi-stage attack scenarios, deployment barriers, and the lack of comparison with commercial automated pentesting platforms.
Specifically, unreliable evaluation can understate the true difficulty of complex attack chains; training-data scarcity limits the strategies that agents can discover; and the absence of robust safety mechanisms blocks real-world deployment, which in turn limits access to the diverse operational data needed to improve training. The lack of commercial baselines further compounds these issues: research prototypes are rarely evaluated alongside deployed automated pentesting products, making it difficult to ground reported progress against real-world capability.

\noindent
$\bullet$
The most foundational challenge is evaluation reliability. Binary task-completion metrics, which dominate across benchmark categories in \S\ref{sec:benchmarks}, cannot distinguish whether an agent succeeds through genuine reasoning, benchmark familiarity, or alignment with the toolchain and environment used during system design. The performance gap between purpose-built benchmarks and shared benchmarks such as AutoPenBench~\cite{gioacchini2025autopenbench} shows that reported success rates are highly sensitive to evaluation setting, making it difficult to compare system designs reliably. A shared benchmark that covers multiple attack classes, enforces clean train–test separation, and reports partial-credit metrics for individual subtasks would enable more reliable cross-system comparison.

\noindent
$\bullet$
The second challenge is the technical ceiling of current systems on realistic multi-host engagements. Multi-agent coordination systems have substantially improved task-completion rates on shared benchmarks, but their gains remain concentrated on short or well-scoped engagements. Long-horizon attack chains that span reconnaissance, initial access, lateral movement, and post-exploitation in realistic enterprise networks remain largely unsolved. These scenarios combine long-context requirements, sparse successful training trajectories, and costly environment resets, placing them beyond the practical reach of current RLVR training pipelines. An open question is whether larger episode budgets and better exploration strategies will be sufficient, or whether fundamentally new training architectures are required.

\noindent
$\bullet$
The third challenge concerns the deployment barriers. Happe~\cite{happe2025surprising} identifies six barriers to deployment, including reliability, safety alignment, privacy sovereignty, ecological cost, accountability, and legal uncertainty, none of which is fully addressed by current system designs. Among these, safety alignment and legal accountability are the most urgent: autonomous agents that execute offensive actions without bounded scope can create direct legal liability for operators, yet current systems provide no formal guarantee that an engagement will remain within the authorized perimeter. These barriers are especially acute in services computing deployments, where an autonomous agent must operate across multi-tenant cloud boundaries, respect authorization contracts between service providers and consumers~\cite{liu2015fine}, and produce audit logs compatible with governance requirements that cloud service operators are legally required to maintain~\cite{ray2013trust}. The Intelligent Assurance System surveyed in \S\ref{sec:defense} represents an initial step toward accountability-aware Agent4Pentest design, but the remaining barriers require coordinated progress across technical, regulatory, and organizational dimensions.

\noindent
$\bullet$
A fourth challenge is the lack of direct comparison with commercial automated pentesting platforms. Dedicated commercial systems such as Pentera~\cite{Pentera2026} and NodeZero~\cite{NodeZero2026} are closed products whose internal architectures and performance figures are not publicly reported, and none of the 81 papers in our corpus evaluates them on a shared benchmark. The recent large-scale empirical study by Peng et al.~\cite{peng2026hackers} addresses this gap only partially, using general-purpose AI coding agents as commercial proxies rather than purpose-built pentesting products. This creates a structural blind spot: the field cannot determine whether research prototypes have surpassed, matched, or still fall short of deployed commercial capability.

Moreover, our corpus includes papers that use commercial LLM APIs such as OpenAI GPT and Anthropic Claude as backbone reasoning engines, but this reflects a deliberate scope decision rather than an inconsistency with the commercial-baseline concern above. In these papers, the research contribution lies in the agent architecture built above the model, including memory management, attack planning, tool integration, and reflection mechanisms, rather than in the underlying model itself. The commercial origin of the backbone model does not make a research prototype equivalent to a commercial pentesting product. Excluding such papers would omit much of the Agent4Pentest literature published between 2023 and 2026. The open challenge is therefore to design an evaluation protocol, anchored by a shared benchmark with clean training-data isolation, that allows commercial products to participate alongside open-source and research-stage systems in a unified capability comparison.

Overall, these non-trivial challenges should be treated as a coupled research agenda rather than as isolated engineering problems. Progress on evaluation reliability, multi-stage capability, and deployment safety depends on progress in the other two dimensions, while the commercial benchmarking gap prevents the field from grounding reported advances against deployed systems and learning which architectural choices matter most in practice. The Agent4Pentest community therefore needs comprehensive evaluation and deployment protocols that jointly address capability measurement, training realism, operational safety, and commercial transparency.
\section{Conclusion}\label{sec:conclusion}

This survey systematically analyzes 81 Agent4Pentest papers between 2023 and 2026. We organize the literature into a six-category taxonomy and trace a four-phase architectural evolution, showing how successive systems address bottlenecks from execution autonomy to sample-efficient RLVR training.
Our analysis shows that Agent4Pentest capability has co-evolved with its evaluation infrastructure. Benchmarks have expanded from CTF platforms to enterprise-scale vulnerability suites, while CTF environments have also become RL training substrates. This coupling accelerates progress but raises reliability concerns, as reported gains may reflect benchmark alignment rather than real-world capability.
We further identify domain-specific frameworks as a distinct category. These systems achieve strong in-domain performance through formal state encodings, domain knowledge bases, constrained action spaces, and specialized verification oracles, but their gains remain narrow and difficult to compare with general-purpose systems under current evaluation practices.
Overall, this survey provides a shared vocabulary and analytical framework for future Agent4Pentest research. Moving the field from controlled benchmarks to realistic security assessments will require more reliable evaluation, stronger multi-stage attack capability, safer deployment mechanisms, and comparisons with commercial baselines. As services computing architectures grow more heterogeneous, Agent4Pentest is likely to become an important component of secure services computing operations.

\bibliographystyle{IEEEtran}
\bibliography{refs}

\end{document}